\documentclass[11pt]{article}
\usepackage{amssymb}
\usepackage{graphics}
\usepackage{epsfig}
\usepackage{a4wide}

\begin{document}

\newcommand{\ppvq}{$pp \to W_H(Z_H) q_- + X$ }
\newcommand{\ppwq}{$pp \to W_H q_- + X$ }
\newcommand{\ppzq}{$pp \to Z_H q_- + X$ }
\newcommand{\qgwq}{$gq \to W_H q'_-$}
\newcommand{\qgvq}{$gq \to W_H(Z_H) q'_-$}
\newcommand{\qgwqx}{$gq \to W_H q'_- + X$}
\newcommand{\qgwqg}{$gq \to W_H q'_- + g$}
\newcommand{\qgwqq}{$gq \to W_H q'_- + q$}
\newcommand{\qqwqq}{$qq' \to W_H q''_- + q'$}
\newcommand{\ppqqwq}{$pp \to gg/q\bar{q} \to q'_- \bar{q}'_H \to W_H q'_- q''$ }
\newcommand{\qqwq}{$gg/q\bar{q} \to q_-' \bar{q}'_H \to W_H q'_- q''$}
\newcommand{\ppwqj}{$pp \to W_H q_- + j$}
\newcommand{\ppvqx}{$pp \to W_H(Z_H) q_- + X$}
\newcommand{\qgvqx}{$gq \to W_H(Z_H) q'_- + X$}
\newcommand{\qgvqg}{$gq \to W_H(Z_H) q'_- + g$}
\newcommand{\qgvqq}{$gq \to W_H(Z_H) q'_- + q$}
\newcommand{\qqvqq}{$qq' \to W_H(Z_H) q''_- + q'$}
\newcommand{\ppqqvq}{$pp \to gg/q\bar{q} \to q'_- \bar{q}'_H \to W_H(Z_H) q'_- q''$ }
\newcommand{\qqvq}{$gg/q\bar{q} \to q_-' \bar{q}'_H \to W_H(Z_H) q'_- q''$}
\newcommand{\ppvqj}{$pp \to W_H(Z_H) q_- + j$}
\newcommand{\etmiss}{\not\hskip-5truedd E_{T} }

\title{$W_H/Z_H$ production associated with a T-odd (anti)quark at the LHC in NLO QCD }
\author{ Zhang Ren-You$^1$, Yan Han$^1$, Ma Wen-Gan$^1$, Wang Shao-Ming$^{1,2}$, Guo Lei$^1$ and Han Liang$^1$ \\
{\small $^1$ Department of Modern Physics, University of Science and Technology}  \\
{\small of China (USTC), Hefei, Anhui 230026, P.R.China} \\
{\small $^2$ Department of Physics, Chongqing University, Chongqing 401331, P.R. China   }}

\date{}
\maketitle \vskip 15mm
\begin{abstract}
In the framework of the littlest Higgs model with T parity, we study
the $W_H/Z_H$ production in association with a T-odd (anti)quark of
the first two generations at the CERN Large Hadron Collider up to
the QCD next-to-leading order. The kinematic distributions of final
decay products and the theoretical dependence of the cross section
on the factorization/renormalization scale are discussed. We apply
three schemes in considering the QCD NLO contributions and find that
the QCD NLO corrections by adopting the (II) and (III) subtraction
schemes can keep the convergence of the perturbative QCD description
and reduce the scale uncertainty of the leading order cross section.
By using these two subtraction schemes, the QCD NLO corrections to
the $W_H(Z_H) q_-$ production process enhance the leading order
cross section with a K-factor in the range of $1.00 \sim 1.43$.
\end{abstract}


\vfill
\eject
\baselineskip=0.32in

\renewcommand{\theequation}{\arabic{section}.\arabic{equation}}
\renewcommand{\thesection}{\Roman{section}.}
\newcommand{\nb}{\nonumber}

\newcommand{\Dir}{\kern -6.4pt\Big{/}}
\newcommand{\Dirin}{\kern -10.4pt\Big{/}\kern 4.4pt}
\newcommand{\DDir}{\kern -7.6pt\Big{/}}
\newcommand{\DGir}{\kern -6.0pt\Big{/}}

\makeatletter      
\@addtoreset{equation}{section}
\makeatother       

\par
\section{Introduction}
\par
The standard model (SM) \cite{s1,s2} provides a remarkably
successful description of high energy physics at the energy scale up
to $10^2~GeV$. Despite its tremendous success, the mechanism of
electroweak symmetry breaking (EWSB) remains the most prominent
mystery in current particle physics, and the smallness of the Higgs
boson mass cannot be protected against perturbative quantum
corrections. The instability of the Higgs boson mass leads to the
so-called "hierarchy problem" between the electroweak scale, $m_H$,
and the cutoff scale of the SM, $\Lambda \sim 10~TeV$
\cite{Barbieri:2000gf}. However, the cutoff scale larger than
$10~TeV$ will lead to a large radiative correction to the Higgs
boson mass, which needs unnatural fine-tuning to give a proper EWSB
scale. To solve the fine-tuning problem (or hierarchy problem) has
become one of the main motivations to the construction of physics
beyond the SM. Besides the supersymmetric and extra dimensions
models, the little Higgs models \cite{LittleHiggs} are very
attractive theories which offer an alternative approach to solve the
"hierarchy problem" \cite{LittleHiggs,Arkani}, and are proposed as
one kind of EWSB models without fine-tuning in which the Higgs boson
is naturally light as a result of nonlinearly realized symmetry
\cite{Arkani}-\cite{LH7}. The simplest version of the little Higgs
models is the littlest Higgs (LH) model \cite{LH4}, which is based
on an $SU(5)/SO(5)$ nonlinear $\sigma$ model \cite{LH5}. However,
precision electroweak constraints \cite{EWC} require the LH model to
characterize a large value of $f$, so the fine-tuning between the
cutoff scale and the electroweak scale is again needed. This problem
can be solved by the littlest Higgs model with T parity (LHT)
\cite{Low:2004xc}-\cite{Cheng:2003ju}.

\par
In the LHT, the SM particles are T-even and their T parity partners
are T-odd. Then the SM gauge bosons cannot mix with the new gauge
bosons, and the electroweak precision observables are not modified
at tree-level. The loop suppression of the corrections to low energy
electroweak observables allows the symmetry breaking scale $f$ to be
significantly lower than $1~TeV$ \cite{Hubisz:2005tx}. In order to
cancel the quadratic divergence of the Higgs boson mass contributed
by top loops, an additional T-even heavy top-quark $T_+$ is
introduced. Then a T-odd partner $T_-$ is also required to implement
the T parity. The CERN Large Hadron Collider (LHC) provides an
opportunity for searching for the new particles predicted in the
LHT. Many studies on the phenomenology of the LHT have been
presented in detail
\cite{Hubisz:2004ft,cpyuan:2006ph,sasha_pheno,Chen:2006ie}. The
phenomenology of these T-odd gauge bosons and fermions in the LHT is
very attractive. Due to the T parity conservation, the T-odd
particles could be produced in pair at the LHC, such as (1) T-odd
fermion pair (or gauge boson pair) production, (2) single T-odd
fermion production associated with a T-odd gauge boson. Of all the
processes in the second type at the LHC, the $W_H q_-$ associated
production, where $q_-=u_-,d_-,c_-, s_-, \bar{u}_-, \bar{d}_-,
\bar{c}_-,\bar{s}_-$, has the largest cross section. Therefore, it
deserves special attention. The production rate of the $pp
\rightarrow Z_H q_- + X$ process is of the same order as, but
quantitatively smaller than, that of the $W_H q_-$ production at the
LHC. The $W_H q_-$ and $Z_H q_-$ production signals could be
detected by their subsequent decays $q_- \to W_H q^{\prime}$, $W_H
\to A_H W$ and $Z_H \to A_H H$, where $q'$ represents $u,d,c,s,
\bar{u}, \bar{d},\bar{c},\bar{s}$ and $A_H$ is the T parity partner
of photon being undetectable as a candidate of dark matter
\cite{Asano}. The phenomenology of the T-odd $SU(2)$ doublet
particle productions at the LHC has been already studied at the
leading-order (LO) \cite{cpyuan:2006ph}. As we know, the LO
predictions for the processes at hadron colliders are always
sensitive to the factorization and renormalization scales ($\mu_f$
and $\mu_r$). For the LO $W_H(Z_H)q_{-}$ production at the LHC
$\mu_f$ enters solely through the parton distribution functions
(PDFs), while the parton level cross section does not depend on
$\mu_r$ at this order. In general, the high order contributions can
reduce the scale uncertainty of the LO cross section, and the QCD
next-to-leading order (NLO) corrections enhance the LO cross
section. Therefore, it is important to take into account the QCD NLO
corrections to reduce the sensitivity to these scales.

\par
In this paper, we focus on the $W_H/Z_H$ production in association
with a T-odd (anti)quark of the first two generations at the LHC,
$pp \to W_H(Z_H) q_- + X$ $(q_-=u_-,d_-,c_-, s_-, \bar{u}_-,
\bar{d}_-, \bar{c}_-,\bar{s}_-)$, up to the QCD NLO. The paper is
organized as follows. In Sec.II, we briefly review the relevant
masses and couplings in the LHT. The detailed strategies of
calculation are given in Sec.III. The numerical results and
discussions are presented in Sec.IV. Finally we present a short
summary.

\vskip 5mm
\section{Related LHT theory}\label{calc}
\par
Based on an $SU(5)/SO(5)$ global symmetry breaking pattern, a
subgroup $[SU(2)\times U(1)]_{1}\times[SU(2)\times U(1)]_{2}$ of the
$SU(5)$ global symmetry is gauged in the LH model
\cite{Cheng:2003ju}, and the gauge fields $W_{i \mu}^a$ and $B_{i
\mu}$ $(a = 1, 2, 3,~ i = 1, 2)$ are introduced. The kinetic terms
for the gauge and scalar fields can be written as
\begin{eqnarray}
 {\cal L}_{G+S} = \sum_{j=1}^2 \left[
 -\frac{1}{2} {\rm Tr} \Big( W_{j\mu\nu}W_j^{\mu\nu} \Big) -
 \frac{1}{4} B_{j\mu\nu}B_j^{\mu\nu} \right] + \frac{f^2}{8} {\rm Tr}
 \left[ \Big( D_{\mu}\Sigma \Big)^{\dag} \Big( D^{\mu} \Sigma \Big) \right],
\end{eqnarray}
where $\Sigma$ is the nonlinear $\sigma$ model field of the LH model and
the covariant derivative $D_{\mu} \Sigma$ is defined as
\begin{eqnarray}
 D_{\mu} \Sigma = \partial_{\mu} \Sigma - i \sum_{j=1}^2 \left[ g_j \Big(
  W_{j \mu} \Sigma + \Sigma W_{j \mu}^{T} \Big) + g_j^{\prime} B_{j \mu}
 \left( Y_j \Sigma + \Sigma Y_j \right) \right].
\end{eqnarray}
To implement T parity in the LHT, we make the following T parity
assignment:
\begin{eqnarray}
 && W_{1\mu}^a \longleftrightarrow W_{2\mu}^a,~~~~
 B_{1\mu} \longleftrightarrow B_{2\mu}, \nonumber \\
 && \Pi \longrightarrow -\Omega \Pi \Omega,~~~~
 {\rm where}~~ \Omega = {\rm diag}(1,1,-1,1,1).
\end{eqnarray}
The invariance of the above Lagrangian under T parity implies that
the gauge couplings of the two $SU(2) \times U(1)$ gauge groups have
to be equal, i.e., $g_1 = g_2 = \sqrt{2} g$, $g_1^{\prime} =
g_2^{\prime} = \sqrt{2} g^{\prime}$.

\par
The gauge symmetry $[SU(2) \times U(1)]_1 \times [SU(2) \times
U(1)]_2$ breaks down to its diagonal subgroup $SU(2) \times U(1)$,
which is generated by the combinations $\{ (Q_1^a+Q_2^a)/\sqrt{2},~
Y_1 + Y_2 \}$, where $\{ Q^{a}_i, Y_i \}$ $(a = 1, 2, 3,~ i = 1, 2)$
are the generators of the two $SU(2) \times U(1)$ gauge groups
respectively. This subgroup is identified with the SM electroweak
gauge group $SU(2)_L \times U(1)_Y$. The corresponding gauge fields
of the residual gauge symmetry, $SU(2)_L \times U(1)_Y$, are just
the T-even eigenstates of the gauge sector:
\begin{eqnarray}
 W_L^a = \frac{W_1^a + W_2^a}{\sqrt{2}}~~~~{\rm and}~~~~
 B_L = \frac{B_1 + B_2}{\sqrt{2}}.
\end{eqnarray}
In addition, the other four orthogonal linear combinations of gauge
fields,
\begin{eqnarray}
 W_H^a = \frac{W_1^a - W_2^a}{\sqrt{2}}~~~~{\rm and}~~~~
 B_H = \frac{B_1 - B_2}{\sqrt{2}},
\end{eqnarray}
are odd under T parity.

\par
The mass eigenstates of the gauge sector in the LHT are expressed as
\begin{eqnarray}
&& W_L^{\pm} = \frac{W_L^1 \mp i W_L^2}{\sqrt{2}},~~~~
  \left( \begin{array}{c} A_L \\ Z_L \end{array} \right)
 = \left( \begin{array}{rc} \cos\theta_W & \sin\theta_W \\
 -\sin\theta_W & \cos\theta_W \end{array} \right)  \left(
 \begin{array}{c} B_L \\ W_L^3 \end{array} \right),~~~~({\rm T-even}), \nonumber \\
&& W_H^{\pm} = \frac{W_H^1 \mp i W_H^2}{\sqrt{2}}, ~~~
  \left( \begin{array}{c} A_H \\ Z_H \end{array} \right)
 = \left( \begin{array}{cr} \cos\theta_H & -\sin\theta_H \\
 \sin\theta_H & \cos\theta_H \end{array} \right)
 \left( \begin{array}{c} B_H \\ W_H^3 \end{array} \right),~~~~({\rm T-odd}),~~~~~
\end{eqnarray}
where the mixing angle $\theta_H$ at the ${\cal O}(v^2/f^2)$ is defined as
\begin{eqnarray}
S_H=\sin\theta_H \simeq \frac{5 g g^{\prime}}{4(5 g^2 - g^{\prime 2})}
\frac{v^2}{f^2},
\end{eqnarray}
and $v \simeq 246~{\rm GeV}$ is the vacuum expectation value of the SM Higgs.

\par
The T-even gauge bosons $A_L$, $Z_L$ and $W_L$ are identified with
the SM photon, $Z$-boson and $W$-boson,
respectively. The masses of the T-odd gauge bosons are given by
\begin{eqnarray}\label{mass-AH-VH}
 m_{A_H} \simeq \frac{1}{\sqrt{5}} g^{\prime} f
 \left( 1 - \frac{5}{8}\frac{v^2}{f^2} \right),~~~~
 m_{W_H} \simeq g f \left( 1 - \frac{1}{8}\frac{v^2}{f^2}
 \right),~~~~ m_{Z_H} \simeq m_{W_H}.
\end{eqnarray}

\par
To implement T parity in the fermion sector (Here we only
present the description of the quark sector as a representative.),
we introduce two incomplete $SU(5)$ multiplets and an $SO(5)$
multiplet:
\begin{eqnarray}
 \Psi_1 = \left( \begin{array}{c} \psi_1 \\ 0 \\ 0 \end{array}
 \right),~~~~  \Psi_2 = \left( \begin{array}{c} 0 \\ 0 \\ \psi_2 \end{array}
 \right),~~~~  \Psi_{HR} = \left( \begin{array}{c} \tilde{\psi}_{HR} \\
 \chi_{HR} \\ \psi_{HR} \end{array} \right), \nonumber \\
 \psi_i = -\tau^2 q_i = -\tau^2 (u_i,~ d_i)^T,~~~~(i = 1, 2, HR),
\end{eqnarray}
which transform under T parity as $\Psi_1 \longrightarrow -\Sigma_0 \Psi_2$,
$\Psi_2 \longrightarrow -\Sigma_0 \Psi_1$ and $\Psi_{HR} \longrightarrow -\Psi_{HR}$,
where $\tau^i~(i=1,2,3)$ are Pauli matrices and
$\Sigma_0$ is a $5 \times 5$ symmetric tensor defined as
\begin{eqnarray}
\Sigma_0 = \langle \Sigma \rangle = \left(
 \begin{array}{ccccc}  & & 1_{2 \times 2} \\  & 1 & \\  1_{2 \times 2} & & \end{array}
 \right).
\end{eqnarray}

\par
The transformations for $\Psi_1$, $\Psi_2$ and $\Psi_{HR}$ under
$SU(5)$ are as $\Psi_1 \longrightarrow V^{*} \Psi_1$, $\Psi_2
\longrightarrow V \Psi_2$ and $\Psi_{HR} \longrightarrow U
\Psi_{HR}$. There $V \in SU(5)$ and $U$ is an $SO(5)$ transformation
in a nonlinear representation of $SU(5)$ defined as
\begin{eqnarray}
 \xi \longrightarrow V \xi U^{\dag} = U \xi \Sigma_0 V^T \Sigma_0~~~
 ({{\rm under~ the}~ SU(5)~ {\rm transformation}~ V}).
\end{eqnarray}
It tells us that $q_1$, $q_2$ and $q_{HR}$ are all $SU(2)$ doublets.
Therefore, we obtain two T-odd $SU(2)$ doublets, $q_H = (q_1 +
q_2)/\sqrt{2}$ and $q_{HR}$, which are left- and right-handed
respectively, and a T-even left-handed $SU(2)$ doublet, $q_{SM} =
(q_1 - q_2)/\sqrt{2}$.

\par
Through the Lagrangian
\begin{eqnarray}
 {\cal L}_F = -\kappa f \Big( \bar{\Psi}_2 \xi +
 \bar{\Psi}_1 \Sigma_0 \Omega \xi^{\dag} \Omega \Big)
 \Psi_{HR} ~+ ~{\rm h.c.},
\end{eqnarray}
the T-odd Dirac fermion doublet $q_-$, defined as $\left( q_-
\right)_L = q_H$ and $\left( q_- \right)_R = q_{HR}$, gains a mass
equal to $\sqrt{2} \kappa f$ before EWSB. After EWSB, a small mass
splitting between the T-odd up- and down-type quarks is induced, and
the masses are given by \cite{Hubisz:2004ft}-\cite{Hubisz:2006mass
splitting}
\begin{eqnarray}\label{m_Q}
 m_{u_-} \simeq \sqrt{2} \kappa f
 \left(
 1 - \frac{1}{8}\frac{v^2}{f^2}
 \right),~~~~
 m_{d_-} = \sqrt{2} \kappa f.
\end{eqnarray}
The T-even left-handed $SU(2)$ doublet $q_{SM}$ is identified with
the left-handed SM fermion doublet. It can acquire Dirac masses
$m_u$ and $m_d$ via Yukawa interactions with the T-even right-handed
$SU(2)$ singlets $u_R$ and $d_R$, respectively. As we know, the
up-type quark is heavier than the down-type quark for each
generation in the SM, while the partners of the SM quarks in a new
physics model may exhibit an inverted mass hierarchy. For example,
the bottom-squarks are considered to be heavier than the top-squarks
in the MSSM, a minimal supersymmetric extension of the SM. As we
expected, Eq.(\ref{m_Q}) indicates that the mass of the T-odd
down-type quark is larger than that of the T-odd up-type quark in
the LHT.

\par
In order to cancel the large quadratic divergent corrections to the
Higgs boson mass induced by the top quark, the Yukawa interaction for
the top sector must be modified. The
$\Psi_1$ and $\Psi_2$ multiplets for the top sector must be
completed to representations of the $SU(3)_1$ and $SU(3)_2$
subgroups of $SU(5)$ by introducing two additional left-handed
$SU(2)$ singlets $U_{L1}$ and $U_{L2}$. These multiplets are
\begin{eqnarray}
 Q_1 = \left( \begin{array}{c} \psi_1 \\ U_{L1} \\ 0
 \end{array}
 \right)~~~{\rm and}~~~ Q_2 =
 \left( \begin{array}{c} 0 \\ U_{L2} \\ \psi_2
 \end{array} \right),
\end{eqnarray}
which obey the same transformation laws under T parity and $SU(5)$
as do $\Psi_1$ and $\Psi_2$. In addition to the T-even right-handed
$SU(2)$ singlet $u_R$, the top sector contains two right-handed
$SU(2)$ singlets $U_{R1}$ and $U_{R2}$, which transform under T
parity as $U_{R1} \longleftrightarrow -U_{R2}$. By using these new
$SU(2)$ singlets introduced in the top sector, we obtain four
additional T parity eigenstates:
\begin{eqnarray}
 U_{L \pm} = \frac{U_{L1} \mp U_{L2}}{\sqrt{2}}~~~{\rm and}~~~
 U_{R \pm} = \frac{U_{R1} \mp U_{R2}}{\sqrt{2}}.
\end{eqnarray}
Therefore, two new heavy partners with opposite T parity, $T_+$ and
$T_-$, should appear in the top sector.

\par
As $U_{L-}$ and $U_{R-}$ do not mix with $u_H$ and $u_{HR}$,
where $u_H$ is the up component of the $SU(2)$ doublet $q_H$,
$T_-$ is simply given by $\left( T_- \right)_L = U_{L-}$
and $\left( T_- \right)_R = U_{R-}$.
However, the T-even eigenstates $U_{L+}$ and $U_{R+}$ mix with
$u_{SM}$ and $u_R$ respectively,
where $u_{SM}$ is the up component of the $SU(2)$ doublet $q_{SM}$,
so that the mass eigenstates of the top quark $t$
and its heavy partner $T_+$ are given by
\begin{eqnarray}
&& \left( \begin{array}{c} t_L \\
 \left(T_+\right)_L \end{array} \right) =
 \left( \begin{array}{cr} \cos\theta_L & -\sin\theta_L \\
 \sin\theta_L & \cos\theta_L \end{array} \right) \left(
 \begin{array}{c} u_{SM} \\ U_{L+} \end{array} \right),   \nb \\
&& \left( \begin{array}{c} t_R \\ \left(T_+\right)_R \end{array}
 \right) = \left( \begin{array}{cr} \cos\theta_R & -\sin\theta_R \\
 \sin\theta_R & \cos\theta_R \end{array} \right) \left( \begin{array}{c}
 u_{R} \\ U_{R+} \end{array} \right),
\end{eqnarray}
where the mixing angles $\theta_{L,R}$ and masses of $T_{\pm}$ are
determined by the Yukawa interaction Lagrangian for the top sector.

\par
The couplings of the T-odd $SU(2)$ doublet quarks and gauge bosons
to the T-even SM particles used in our calculations are listed in
Table \ref{tab1} \cite{Hubisz:2004ft,Blanke:2007ckm}, where
$(V_{Hu})_{ij}$ and $(V_{Hd})_{ij}$ are the matrix elements of the
CKM-like unitary mixing matrices $V_{Hu}$ and $V_{Hd}$,
respectively. The two mixing matrices satisfy
$V_{Hu}^{\dag}V_{Hd}=V_{CKM}$ \cite{Blanke:2007ckm}, therefore, they
cannot simultaneously be set to the identity. In the following
calculations we take $V_{Hu}$ to be a unit matrix, then we have
$V_{Hd}=V_{CKM}$.

\begin{table}[h]
\begin{center}
\begin{tabular}{|c|l||c|l|}
\hline
Interaction & ~~~~~~~~Feynman rule & Interaction & ~~~~~~~~~Feynman rule \\
\hline
&&& \\
$W_{H}^{+\mu} \bar{u}_-^i d^j$ &
$i\frac{g}{\sqrt{2}}(V_{Hd})_{ij}\gamma^\mu  P_L$ &
$W_{H}^{-\mu} \bar{d^i}_- u^j$ & $i\frac{g}{\sqrt{2}}(V_{Hu})_{ij}\gamma^\mu P_L$\\
&&& \\
$Z_{H}^{\mu} \bar{u}_-^i u^j$ & $i(\frac{g C_H}{2}-\frac{g'
S_H}{10})(V_{Hu})_{ij} \gamma^\mu P_L$ &
$Z_{H}^{\mu} \bar{d}_-^i d^j$ & $i(-\frac{g C_H}{2}-\frac{g' S_H}{10})(V_{Hd})_{ij} \gamma^\mu P_L$ \\
&&& \\
$\bar{q}_{-}^{\alpha} q_{-}^{\beta} G^{a}_{\mu}$ & $ig_s (T^a)_{\alpha\beta}\gamma^{\mu}$ && \\
&&& \\
\hline
\end{tabular}
\caption{\label{tab1} The related LHT Feynman rules used in this
work, where $q_-=u_-,d_-,c_-,s_-,t_-,b_-$, $i$ and $j$ are the
generation indices and $C_H^2=1-S_H^2$.}
\end{center}
\end{table}

\par
\section{Analytic calculations}
\subsection{LO cross sections }
\par
\begin{figure}
\begin{center}
\includegraphics[width=0.44\textwidth]{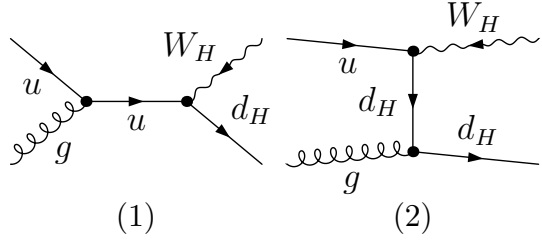}
\caption{ \label{fig1} The LO Feynman diagrams for the partonic
process $gu\to W^+_H d_-$. }
\end{center}
\end{figure}

The partonic processes, which contribute to the parent process \ppvq
at the LHC, are written as
\begin{eqnarray}
\label{process}
g(p_{1})+ q(p_{2})\to V_H(p_{3})+q_-'(p_{4}), &&
(V_H=W_H,Z_H,~q=u,d,c,s,\bar u,\bar d,\bar c,\bar s),  \nb \\
&& (q_-'=u_-,d_-,c_-,s_-,\bar u_-,\bar d_-,\bar c_-,\bar s_-).
\end{eqnarray}
There are two LO Feynman diagrams for each of the above partonic
processes. We plot the LO Feynman diagrams for the partonic process
$gu\to W^+_H d_-$ as a representative in Fig.\ref{fig1}. The
Fig.\ref{fig1}(1) and Fig.\ref{fig1}(2) diagrams are $s$- and
$t$-channel Feynman diagrams respectively. The LO cross section for
the partonic process $gq \rightarrow V_H q_-^{\prime}$ has the form
as
\begin{eqnarray}
\hat{\sigma}_{LO}(\hat{s}, gq \to V_H q_-')= \frac{(2 \pi
)^4}{4|\vec{p}_1|\sqrt{\hat{s}}}\int \overline{\sum} |{\cal
M}_{LO}|^2 d\Phi_2,~~(q=u,d,c,s,\bar u,\bar d,\bar c,\bar s),
\end{eqnarray}
where $d\Phi_2$ is the two-body phase space element, and $\vec{p}_1$
is the momentum of the initial gluon in the center-of-mass system.
The integration is performed over the two-body phase space of the
final particles $V_H$ and $q'_-$. The summation is taken over the
spins and colors of the initial and final states, and the bar over
the summation indicates averaging over the intrinsic degrees of
freedom of initial partons.

\par
The LO total cross section for the parent process $pp \to V_H q_-+X$
can be expressed as
\begin{eqnarray}\label{sigma_PP}
&& \sigma_{LO}(pp \to V_H q_-+X) =    \nb \\
&& \sum_{q=u,d,c,s,}^{\bar{u},\bar{d},\bar{c},\bar{s}} \left\{ \int
dx_A dx_B\left[ G_{g/A}(x_A,\mu_f)
G_{q/B}(x_B,\mu_f)\hat{\sigma}_{LO}(gq \to V_H
q'_-, x_{A}x_{B}s,\mu_f,\mu_r)+(A\leftrightarrow B)\right] \right\}, \nb \\
\end{eqnarray}
where $G_{i/P}$ $(i=g,q,~P=A,B)$ represents the PDF of parton $i$ in
proton $P$, $x_P$ $(P=A,B)$ is the momentum fraction of a parton
(gluon or quark) in proton $P$, and $\mu_f$ and $\mu_r$ are the
factorization and renormalization scales, respectively.

\par
\subsection{QCD NLO corrections } \label{Sec II.2}
\par

\par
\subsubsection{General description} \label{Sec II.2.1}
\par
The QCD NLO corrections to the \ppvq process involve the following
components:
\begin{itemize}
 \item[(i)] The QCD one-loop virtual corrections to the partonic
       processes \qgvq.
 \item[(ii)] The contributions of the real gluon emission partonic
       processes \qgvqg.
 \item[(iii)] The contributions of the real light-quark emission partonic processes
       $gg \rightarrow W_H(Z_H) q_-^{\prime} + \bar{q}$,
       $q^{\prime \prime} \bar{q}^{\prime \prime} \rightarrow W_H(Z_H) q_-^{\prime} + \bar{q}$
       and $q q^{\prime \prime} \rightarrow W_H(Z_H) q_-^{\prime} + q^{\prime \prime}$.
 \item[(iv)] The corresponding contributions of the PDF counterterms.
\end{itemize}

\par
\begin{figure}
\begin{center}
\includegraphics[width=0.87\textwidth]{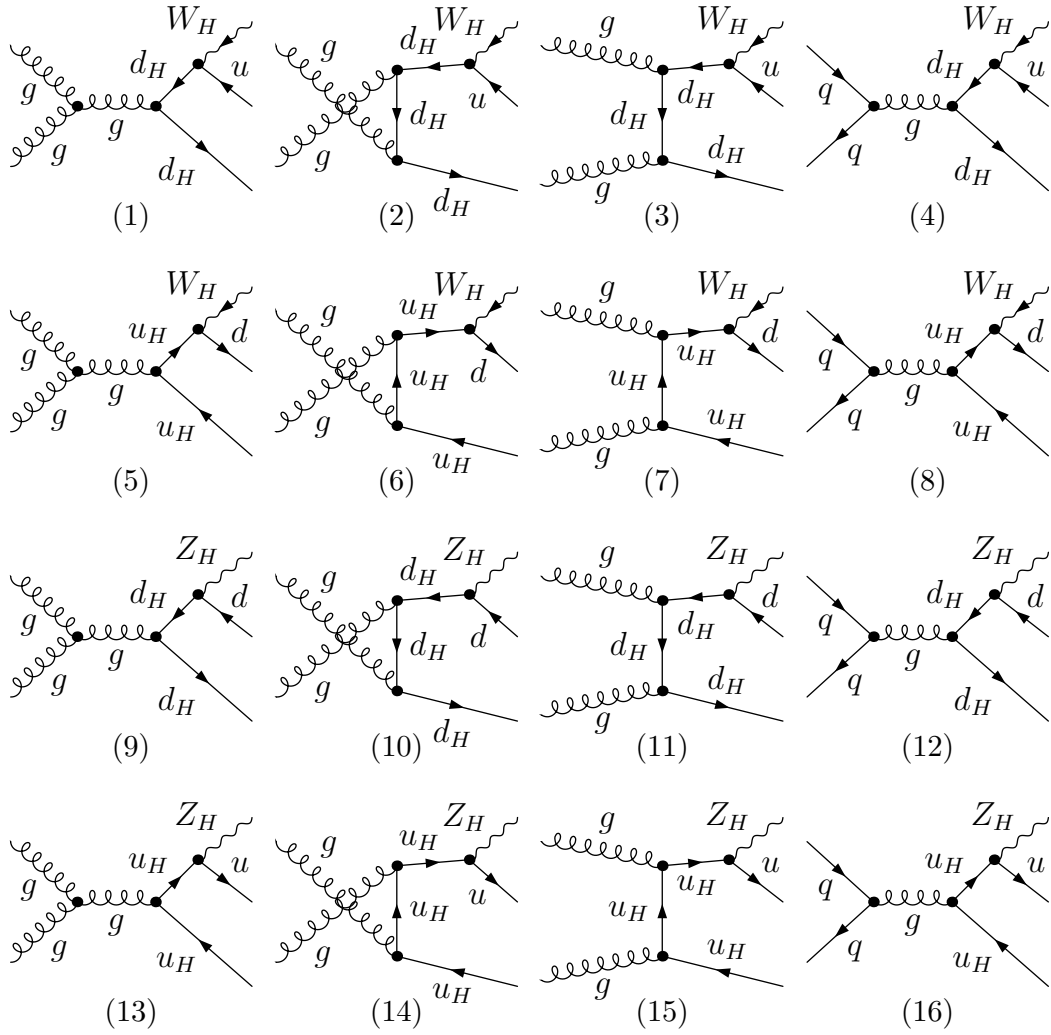}
\caption{ \label{fig2} The Feynman diagrams for the real light-quark
emission partonic processes via intermediate on-shell T-odd quarks.
}
\end{center}
\end{figure}

It should be noticed that for the $gg \rightarrow W_H(Z_H)
q_-^{\prime} + \bar{q}$ and $q^{\prime \prime} \bar{q}^{\prime
\prime} \rightarrow W_H(Z_H) q_-^{\prime} + \bar{q}$ light-quark
emission partonic processes there exists resonance effect due to the
$q_-$ propagator. In Fig.\ref{fig2} we present the Feynman diagrams
for these real light-quark emission partonic processes via
intermediate on-shell T-odd quarks. To deal with the resonance
effect in these partonic processes, we replace $m_{q_-}^2$ in the
denominator of the $q_-$ propagator by $m_{q_-}^2-i
m_{q_-}\Gamma_{q_-}$. With the LHT parameter values used in this
paper, the main decay channels of $q_-$ are $q_- \rightarrow W_H
q^{\prime}$, $q_- \rightarrow Z_H q$ and $q_- \rightarrow A_H q$:
\begin{eqnarray}
{\rm Br}(q_- \to W_H q^{\prime}) + {\rm Br}(q_- \to Z_H q) + {\rm
Br}(q_- \to A_H q) \simeq 100\%.
\end{eqnarray}
Therefore, the value of $\Gamma_{q_-}$ is obtained approximately by
summing up the LO partial decay widths of these main decay channels.
These QCD NLO contribution parts from the $gg \rightarrow W_H(Z_H)
q_-^{\prime} + \bar{q}$ and $q^{\prime \prime} \bar{q}^{\prime
\prime} \rightarrow W_H(Z_H) q_-^{\prime} + \bar{q}$ partonic
processes are quite large due to the high gluon luminosity and the
$q_-$ resonance effect.

\par
In this work, we apply three schemes in considering the QCD NLO
corrections. In scheme (I) (denoted as "QCD NLO I") we include all
the four components mentioned above in the QCD NLO corrections. With
this scheme, the Feynman diagrams in Fig.\ref{fig2} could lead to
large corrections to the Born $pp \to W_H(Z_H)q_- + X$ process due
to the $q_-$ resonance effect, and destroy the perturbative
convergence. Furthermore, these Feynman diagrams are also counted
towards $q_-\bar{q}_-$ production followed by an on-shell decay $q_-
\to W_H(Z_H) q^{\prime}$. Therefore, to avoid double counting and to
keep the convergence of the perturbative QCD description of the
$W_H(Z_H)q_-$ associated production channel, we should remove the
intermediate on-shell $q_-$ contributions from the $W_H(Z_H) q_-$
associated production \cite{on-shell subtraction}.

\par
In scheme (II) (denoted as "QCD NLO II") we exclude the
contributions of the partonic processes $gg \to W_H(Z_H)
q_-^{\prime} + \bar{q}$ and $q^{\prime \prime} \bar{q}^{\prime
\prime} \to W_H(Z_H) q_-^{\prime} + \bar{q}$ from the QCD NLO
corrections. Since the corrections to the parent process $pp \to
W_H(Z_H)q_- + X$ contributed by the partonic processes $gg
\rightarrow W_H(Z_H) q_-^{\prime} + \bar{q}$ and $q^{\prime \prime}
\bar{q}^{\prime \prime} \to W_H(Z_H) q_-^{\prime} + \bar{q}$ and
their corresponding PDF counterterms are IR-safe, we could exclude
them from the QCD NLO corrections to the parent process $pp \to
W_H(Z_H)q_- + X$. With this subtraction scheme, the intermediate
on-shell $q_-$ contributions to the $W_H(Z_H)q_-$ associated
production are removed and the perturbative convergence is kept.
Since all the $gg$ and $q\bar{q}$ initiated contributions are
excluded, this scheme subtracts some genuine QCD NLO contributions.

\par
We adopt another subtraction strategy, the PROSPINO scheme
\cite{PROSPINO-ref,on-shell subtraction}, which removes the on-shell
T-odd quark pair production from the real light-quark emissions $gg
\rightarrow W_H(Z_H) q_-^{\prime} + \bar{q}$ and $q^{\prime \prime}
\bar{q}^{\prime \prime} \rightarrow W_H(Z_H) q_-^{\prime} +
\bar{q}$, to avoid double counting and to not artificially ruin the
convergence of the perturbative QCD description of the $pp
\rightarrow W_H(Z_H)q_- + X$ process. This on-shell subtraction
scheme can provide a reliable production rate since it only
subtracts the squared on-shell amplitudes and does this point by
point over the entire phase space. We call this subtraction scheme
as the scheme (III) in this work, which is defined as a replacement
of the Breit-Wigner propagator \cite{on-shell subtraction}
\begin{eqnarray}
 \frac{|{\cal M}|^2( s_{V_H q} )}{( s_{V_H q} - m_{q_-}^2 )^2
 + m_{q_-}^2 \Gamma_{q_-}^2}
 & \to &
 \frac{|{\cal M}|^2( s_{V_H q} )}{( s_{V_H q} - m_{q_-}^2 )^2
 + m_{q_-}^2 \Gamma_{q_-}^2} \\
 &&-
 \frac{|{\cal M}|^2( m_{q_-}^2 )}{( s_{V_H q} - m_{q_-}^2 )^2
 + m_{q_-}^2 \Gamma_{q_-}^2}
 \Theta( \hat{s} - 4 m_{q_-}^2 )
 \Theta( m_{q_-} - m_{V_H} ), \nonumber
\end{eqnarray}
where $s_{V_H q}$ is the squared momentum flowing through the
intermediate $q_-$ propagator. The results by adopting this scheme
are denoted as "QCD NLO III".

\par
To isolate the UV and IR singularities, we adopt the dimensional
regularization method in $D=4-2 \epsilon$ dimensions. The collinear
counterterm of the PDF, $\delta G_{i/P}(x,\mu_f)$ ($P=$ proton,
$i=g,u,\bar{u}$, $d,\bar{d}$, $c,\bar{c}$, $s,\bar{s}$), is split
into two parts: the collinear gluon emission part $\delta
G_{i/P}^{(gluon)}(x,\mu_f)$ and the collinear light-quark emission
part $\delta G_{i/P}^{(quark)}(x,\mu_f)$,
\begin{eqnarray}\label{PDFcounterterm1}
&& \delta G_{q(g)/P}(x,\mu_f) = \delta G_{q(g)/P}^{(gluon)}(x,\mu_f)
                  +\delta G_{q(g)/P}^{(quark)}(x,\mu_f),
                ~~(q = u, \bar{u}, d, \bar{d}, c, \bar{c}, s, \bar{s} ),
\end{eqnarray}
where
\begin{eqnarray}\label{PDFcounterterm2}
&& \delta G_{q(g)/P}^{(gluon)}(x,\mu_f) =
   \frac{1}{\epsilon} \left[
                      \frac{\alpha_s}{2 \pi}
                      \frac{\Gamma(1 - \epsilon)}{\Gamma(1 - 2 \epsilon)}
                      \left( \frac{4 \pi \mu_r^2}{\mu_f^2} \right)^{\epsilon}
                      \right]
   \int_x^1 \frac{dz}{z} P_{qq(gg)}(z) G_{q(g)/P}(x/z,\mu_f), \nonumber \\
&& \delta G_{q/P}^{(quark)}(x,\mu_f) =
   \frac{1}{\epsilon} \left[
                      \frac{\alpha_s}{2 \pi}
                      \frac{\Gamma(1 - \epsilon)}{\Gamma(1 - 2 \epsilon)}
                      \left( \frac{4 \pi \mu_r^2}{\mu_f^2} \right)^{\epsilon}
                      \right]
   \int_x^1 \frac{dz}{z} P_{qg}(z) G_{g/P}(x/z,\mu_f),  \nonumber \\
&& \delta G_{g/P}^{(quark)}(x,\mu_f) =
   \frac{1}{\epsilon} \left[
                      \frac{\alpha_s}{2 \pi}
                      \frac{\Gamma(1 - \epsilon)}{\Gamma(1 - 2 \epsilon)}
                      \left( \frac{4 \pi \mu_r^2}{\mu_f^2} \right)^{\epsilon}
                      \right]
   \sum_{q=u,\bar{u}}^{d,\bar{d}, c, \bar {c}, s, \bar {s}}
   \int_x^1 \frac{dz}{z} P_{gq}(z) G_{q/P}(x/z,\mu_f).~~~~~
\end{eqnarray}
The explicit expressions for the splitting functions
$P_{ij}(z)~(ij=qq,qg,gq,gg)$ in Eqs.(\ref{PDFcounterterm2}) are
available in Ref.\cite{19}.

\par
\subsubsection{Virtual and real emission corrections to \qgvq}
\par
\begin{figure}
\begin{center}
\includegraphics[width=0.7\textwidth]{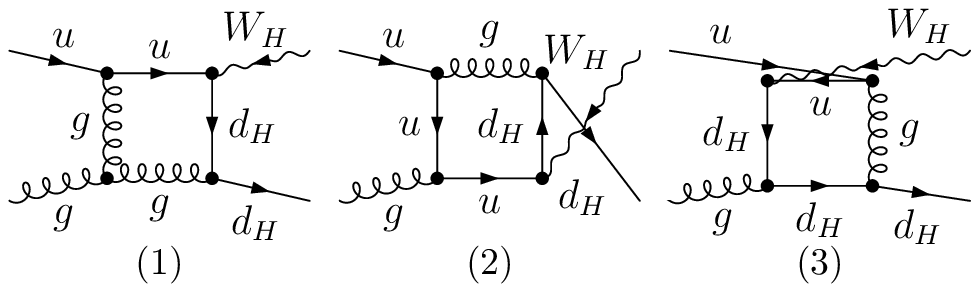}
\caption{ \label{fig3} The box Fynman diagrams for the partonic
process $gu \to W^+_H d_-$. }
\end{center}
\end{figure}

The one-loop level amplitudes for the partonic processes $gq
\rightarrow W_H(Z_H) q_-^{\prime}$ in the LHT include the
contributions of the self-energy, vertex and box graphs. In
Fig.\ref{fig3} the box Feynman diagrams for the partonic process $gu
\to W^+_H d_-$ are presented as a representative.

\par
The strong coupling constant, the masses and wave functions of the
relevant colored particles in the LHT are renormalized to remove the
UV divergences of the virtual corrections.
In the QCD NLO calculations of the $pp \rightarrow W_H(Z_H) q_- + X$
process, the following renormalization constants are introduced:
\begin{eqnarray}
\label{defination of renormalization constants}
\psi_{q(q_-)}^{0,L,R} &=& \left( 1 + \frac{1}{2} \delta
Z_{q(q_-)}^{L,R} \right) \psi_{q(q_-)}^{L,R},~~~~~
m^{0}_{q_-} = m_{q_-} + \delta m_{q_-},~~ \nb \\
G_{\mu}^0
&=&
\left(
1 + \frac{1}{2} \delta Z_g
\right)
G_{\mu},~~~~~~~~~~~~~~
g_s^0 = g_s + \delta g_s,
\end{eqnarray}
where $g_s$ denotes the strong coupling constant, $m_{q_-}$ is the
T-odd quark mass, $\psi_{q(q_-)}^{L,R}$ and $G_{\mu}$ denote the
fields of the SM quark, T-odd heavy quark and gluon, respectively.
The masses and wave functions of the colored fields are renormalized
by adopting the on-shell scheme, then the relevant renormalization
constants are expressed as
\begin{eqnarray} \label{CT-q}
\delta Z_{q}^{L,R}
& \equiv & \delta Z_{q}
=
-\frac{\alpha_s(\mu_r)}{3 \pi}
\Big[
\Delta_{UV}-\Delta_{IR}
\Big],  \\
\delta Z_{q_-}^{L,R}
& \equiv & \delta Z_{q_-}
=
-\frac{\alpha_s(\mu_r)}{3 \pi}
\left[
\Delta_{UV}+2\Delta_{IR}+4+3\ln\left(\frac{\mu_r^2}{m_{q_-}^2}\right)
\right],  \\
\frac{\delta m_{q_-}}{m_{q_-}} &=& - \frac{\alpha_s(\mu_r)}{3 \pi}
\left\{ 3\left[ \Delta_{UV} + \ln\left( \frac{\mu_r^2}{m_{q_-}^2}
\right) \right]+4 \right \},  \\
\label{CT-Zg} \delta Z_g &=& - \frac{\alpha_s(\mu_r)}{2
\pi}\left\{\frac{3}{2}\Delta_{UV}+\frac{5}{6}\Delta_{IR}+
\frac{1}{3}\ln\left(\frac{\mu_r^2}{m_t^2}\right)
+\frac{1}{3}\sum\limits_{T=T_+}^{T_-}\ln\left(\frac{\mu_r^2}{m_{T}^2}\right)
+\frac{1}{3}\sum\limits_{q_-}\ln\frac{\mu_{r}^2} {m_{q_-}^2}\right\}, \nb \\
&&~~~~~~~~~~~~~~~~~~~~~~~~~~~~~~~~~~~~~~~~~~~~~~~~~~~~~~~~~
(q_-=u_-,d_-,c_-,s_-,t_-,b_-),
\end{eqnarray}
where $\Delta_{UV}=1/\epsilon_{UV} -\gamma_E +\ln(4\pi)$ and
$\Delta_{IR}=1/\epsilon_{IR} -\gamma_E +\ln(4\pi)$.

\par
For the renormalization of the strong coupling constant $g_{s}$, we
adopt the $\overline{MS}$ scheme at the renormalization scale
$\mu_{r}$, except that the divergences associated with the massive
top-quark, T-odd $SU(2)$ doublet quarks ($u_-, d_-, c_-, s_-, t_-,
b_-$) and $T_{\pm}$ loops are subtracted at zero momentum \cite{gs}.
Then the renormalization constant of the strong coupling constant
can be obtained as
\begin{eqnarray} \label{CT-g}
\frac{\delta g_s}{g_s}&=& -\frac{\alpha_s(\mu_r)}{4\pi}
\left[\frac{3}{2}\Delta_{UV}+\frac{1}{3}\ln\frac{m_{t}^2}
{\mu_{r}^2}+\frac{1}{3}\sum\limits_{T=T_+}^{T_-}\ln\frac{m_{T}^2}
{\mu_{r}^2}+ \frac{1}{3}\sum\limits_{q_-}\ln\frac{m_{q_-}^2}
{\mu_{r}^2}\right],  \nb \\
&& ~~~~~~~~~~~~~~~~~~~~~~~~~~~~~~~~~~~~~
~~~~~~~(q_-=u_-,d_-,c_-,s_-,t_-,b_-).
\end{eqnarray}

\par
The LO amplitude for $gq \to W_H(Z_H) q_-^{\prime}$ can be expressed
as
\begin{eqnarray}
{\cal M}_{LO} = {\cal M}_s + {\cal M}_t,
\end{eqnarray}
where ${\cal M}_s$ and ${\cal M}_t$ are the amplitudes for the
$s$- and $t$-channel Feynman diagrams, respectively. Then the
QCD NLO counterterm amplitude can be written as
\begin{eqnarray} \label{CT-M}
{\cal M}_{CT}= \left( \frac{\delta g_s}{g_s} + \frac{1}{2} \delta
Z_g + \frac{1}{2} \delta Z_q + \frac{1}{2} \delta Z_{q_-^{\prime}}
\right) {\cal M}_{LO} + \delta m_{q_-^{\prime}} {\cal M}_t
 \Big|_{\frac{i}{(\rlap/p_{q_-^{\prime}} - m_{q_-^{\prime}})}
 \to  \frac{i}{(\rlap/p_{q_-^{\prime}} - m_{q_-^{\prime}})^2}},
\end{eqnarray}
where ${\cal M}_t
 \Big|_{\frac{i}{(\rlap/p_{q_-^{\prime}} - m_{q_-^{\prime}})}
 \to \frac{i}{(\rlap/p_{q_-^{\prime}} - m_{q_-^{\prime}})^2}}$
represents the amplitude obtained from the $t$-channel amplitude
${\cal M}_t$ by doing the replacement of
$\frac{i}{(\rlap/p_{q_-^{\prime}} - m_{q_-^{\prime}})} \to
\frac{i}{(\rlap/p_{q_-^{\prime}} - m_{q_-^{\prime}})^2}$, and
$p_{q_-^{\prime}}$ is the four-momentum of the T-odd quark in the
$t$-channel propagator. From Eqs.(\ref{CT-Zg}) and (\ref{CT-g}) we
can see that the terms of
$\sum\limits_{T=T_+}^{T_-}\ln\frac{m_{T}^2} {\mu_{r}^2}$ are exactly
canceled in Eq.(\ref{CT-M}), therefore, the values of $m_{T_{\pm}}$
are unnecessary in our numerical calculations.

\par
We use our developed in-house programs to isolate analytically the
IR singularities of loop integrals and calculate numerically
one-loop integrals based on the {\sc LoopTools-2.4} package
\cite{formloop,ff}, where the analytical expressions for the
IR-singular parts of loop integrals are adopted from
Ref.\cite{Stefan}, and the numerical evaluations of IR-safe
$N$-point ($N\leq4$) integrals are implemented by using the formulas
in Refs.\cite{OneTwoThree,Four,Five}.

\par
We employ the two cutoff phase space slicing (TCPSS) method
\cite{19} to calculate the corrections from the real
gluon/light-quark emission partonic processes. An arbitrary soft
cutoff $\delta_{s}$ separates the real gluon emission phase space
into two regions, the soft gluon region and the hard gluon region.
Another cutoff $\delta_{c}$ decomposes the real hard
gluon/light-quark emission phase space region into the hard
collinear ($HC$) region and the hard noncollinear ($\overline{HC}$)
region. Then the soft and collinear IR singularities are isolated
from the IR-safe region. The integration over the $\overline{HC}$
region of phase space is performed in the four-dimensions by using
the Monte Carlo integrator \cite{Lepage}. Finally, the total cross
section for the real emission process can be expressed as
\begin{equation}
\label{sigmaR} \Delta \sigma_{R}=\Delta \sigma_{S}+\Delta \sigma_{H}
=\Delta \sigma_{S}+\Delta \sigma_{HC}+\Delta \sigma_{\overline{HC}}.
\end{equation}

\par
The UV singularities of the loop corrections are canceled by those
of the related counterterms contributed by the renormalization
constants in Eqs.(\ref{CT-q})-(\ref{CT-g}). Therefore, the
renormalized virtual corrections (loop corrections combined with the
related counterterms) are UV-finite. Furthermore, the renormalized
virtual corrections also contain soft and collinear IR
singularities. These IR singularities exactly vanish after combining
the renormalized virtual corrections with the contributions of the
real gluon/light-quark emission processes and the PDF counterterms.
These cancelations have been verified analytically and numerically
in our calculations.

\vskip 5mm
\section{Numerical results and discussions }\label{numres}
\par
\subsection{Input parameters}\label{parameters}
\par
In the study of the dependence of the QCD NLO corrected cross
section on the factorization and renormalization scales, we set the
two unphysical scales equal to a common value ($\mu_f = \mu_r =
\mu$) and do not vary them in an independent way for simplicity.
This setting of scales may render the results more stable than they
actually are due to the logarithmic term
$\ln\frac{\mu_r^2}{\mu_f^2}$ in the QCD NLO contributions. For
example, the PDF counterterms in Eqs.(\ref{PDFcounterterm2}) have
the form as $\frac{\alpha_s}{2 \pi} \Big( \frac{1}{\epsilon} -
\gamma_E + \ln4 \pi + \ln\frac{\mu_r^2}{\mu_f^2}\Big) ( P \otimes G
)$, where $P \otimes G$ represents the convolution of the splitting
function $P$ with the PDF $G$. When we set $\mu_f = \mu_r$, we
obtain $\ln\frac{\mu_r^2}{\mu_f^2} = 0$ and the
factorization/renormalization scale dependence of these PDF
counterterms is underestimated.

\par
We take one-loop and two-loop running $\alpha_{s}$ in the LO and QCD
NLO calculations, respectively \cite{hepdata}. The central value of
the factorization/renormalization scale $\mu$ is chosen as
$\mu_0=(m_{W_H}+m_{d_-})/2$. We adopt the CTEQ6L1 and CTEQ6M parton
densities with five flavors in the LO and NLO calculations,
respectively \cite{cteq}. The strong coupling constant
$\alpha_s(\mu)$ is determined by the QCD parameter $\Lambda_5^{LO} =
165~MeV$ for the CTEQ6L1 at the LO and $\Lambda_5^{\overline{MS}} =
226~MeV$ for the CTEQ6M at the NLO \cite{hepdata}. We ignore the
masses of $u$-, $d$-, $c$-, $s$-, $b$-quarks, and take
$\alpha_{ew}(m_Z^2)^{-1}|_{\overline{MS}}=127.925$,
$m_W=80.399~GeV$, $m_Z=91.1876~GeV$, $m_t=171.2~GeV$ and
$\sin^2\theta_W=1-\left(\frac{m_W}{m_Z}\right)^2=0.222646$.

\par
The colliding energy in the proton-proton center-of-mass system is
taken as $\sqrt s=7~TeV$ for the early LHC and $\sqrt s=14~TeV$ for
the later running at the LHC. The Cabibbo-Kobayashi-Maskawa (CKM)
matrix elements are taken as
\begin{eqnarray}\label{CKM}
 V_{CKM} &=& \left(
\begin{array}{ccc}
    V_{ud} \ &  V_{us} \ &  V_{ub} \\
    V_{cd} \ &  V_{cs} \ &  V_{cb} \\
    V_{td} \ &  V_{ts} \ &  V_{tb} \\
\end{array}
    \right)=\left(
\begin{array}{ccc}
     0.97418 \ &  0.22577 \ &  0 \\
    -0.22577 \ &  0.97418 \ &  0 \\
       0 \ &  0 \ &  1 \\
\end{array}  \right).
\end{eqnarray}

\par
\subsection{Independence on two cutoffs}
\par
\begin{figure}[htbp]
\begin{center}
\includegraphics[scale=0.5]{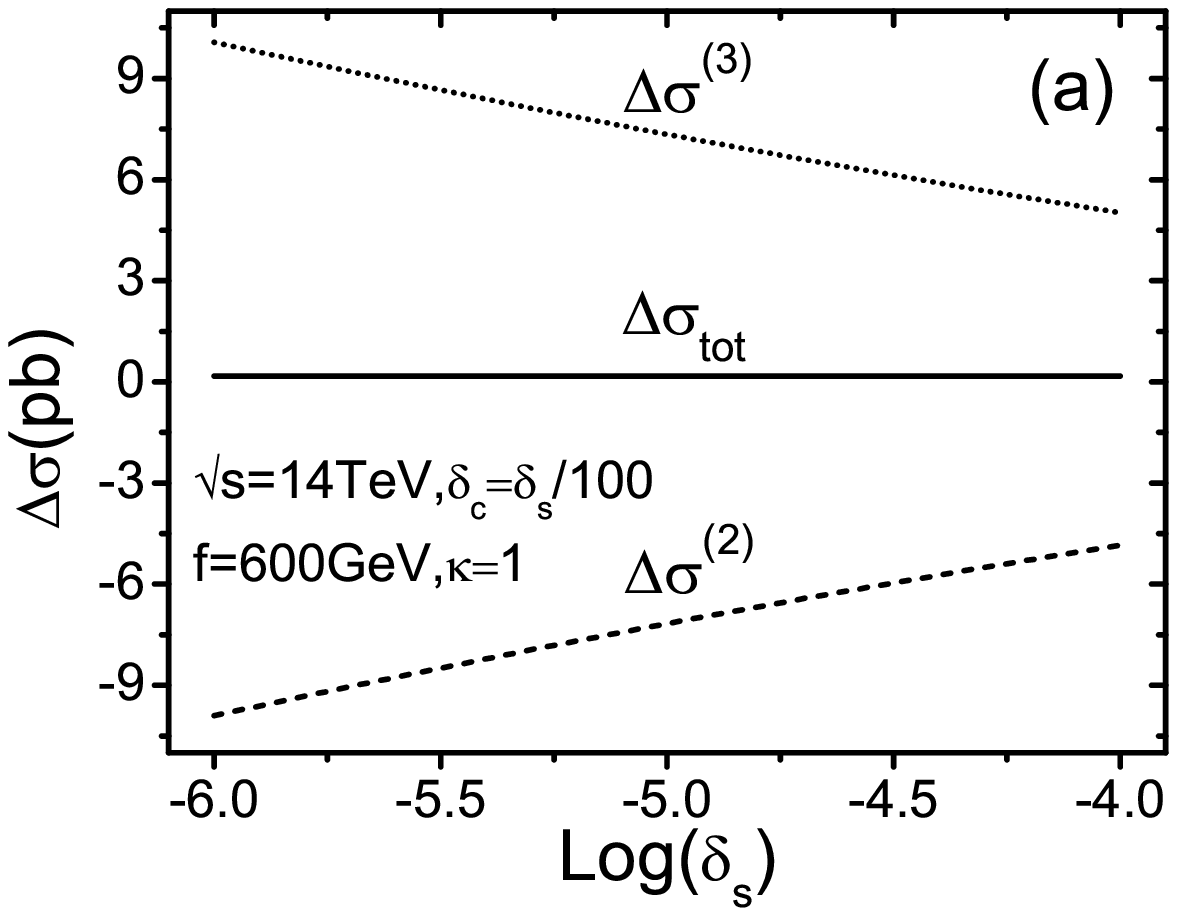}
\includegraphics[scale=0.42]{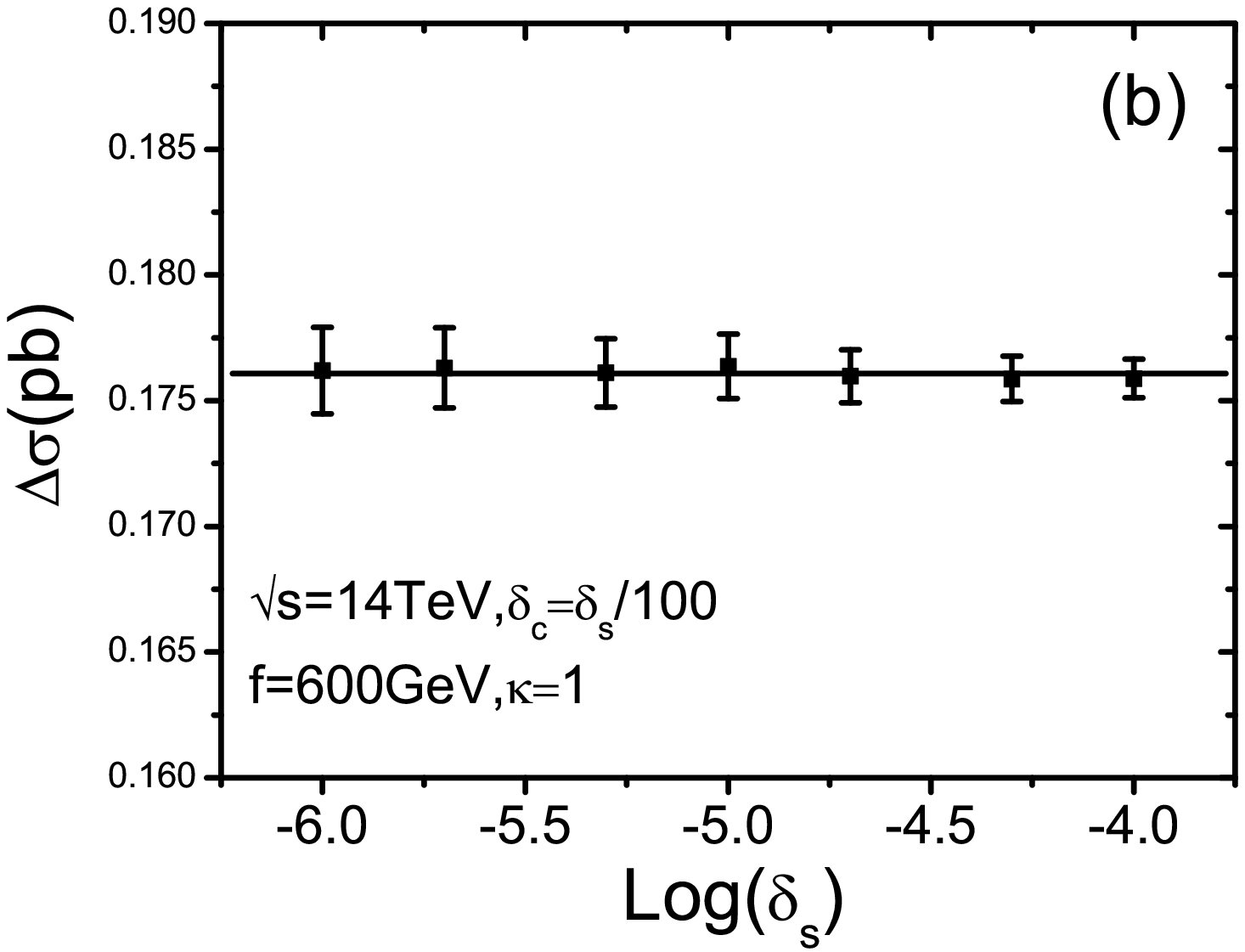}
\hspace{0in}%
\caption{\label{fig4} (a) The dependence of the QCD NLO corrections
to the $pp \to ug \to W_H^+ d_- + X$ process on the cutoffs $\delta_s$
and $\delta_c$ at the LHC, where $f=600~GeV$, $\kappa=1$,
$\delta_c=\delta_s/100$ and
$\mu=\mu_0=(m_{W_H}+m_{d_-})/2=623.55~GeV$. (b) The amplified curve
for $\Delta\sigma_{tot}$ of Fig.\ref{fig4}(a). }
\end{center}
\end{figure}

In order to demonstrate the independence of the total QCD NLO
correction to the $pp \to ug \to W_H^+ d_- + X$ process on the two
cutoffs, $\delta_s$ and $\delta_c$, we present the QCD NLO
correction parts as the functions of the cutoffs in
Fig.\ref{fig4}(a), where we take $\delta_c=\delta_s/100$,
$f=600~GeV$ and $\kappa=1$. From Eqs.(\ref{mass-AH-VH}) and
(\ref{m_Q}), we obtain the related masses of the T-odd particles as
$m_{W_H}=398.57~GeV$, $m_{u_-}=m_{c_-}=830.70~GeV$ and
$m_{d_-}=m_{s_-}=848.53~GeV$. In this figure we take
$\mu=\mu_0\equiv (m_{W_H}+m_{d_-})/2=623.55~GeV$. Although the decay
width of $q_-$ is less than $1 \%$ of $m_{q_-}$, we take
$\Gamma_{d_-} = 0.1m_{d_-}$ to suppress the resonance effect, which
makes the cutoff independence more clear. The amplified curve for
$\Delta\sigma_{tot}$ of Fig.\ref{fig4}(a) is shown in
Fig.\ref{fig4}(b). The figures demonstrate that the total QCD NLO
correction $\Delta\sigma_{tot}$ which is the summation of the
two-body and three-body corrections, is independent of the two
cutoffs within the statistical errors, even though the two-body
correction ($\Delta\sigma^{(2)}$) and three-body correction
($\Delta\sigma^{(3)}$) are strongly influenced by the cutoffs
$\delta_s$ and $\delta_c$. As we know, the independence of the total
QCD NLO correction to the $pp \rightarrow ug \rightarrow W_H^+
d_-+X$ process on the cutoffs $\delta_s$ and $\delta_c$ is a
necessary condition that must be fulfilled for the correctness of
our calculations. In the further numerical calculations, we fix
$\delta_s=1 \times 10^{-4}$ and $\delta_c=\delta_s/100$.

\par
\subsection{Dependence on factorization/renormalization scale \label{sub-0} }
\par
\begin{figure}[htbp]
\begin{center}
\includegraphics[width=0.45\textwidth]{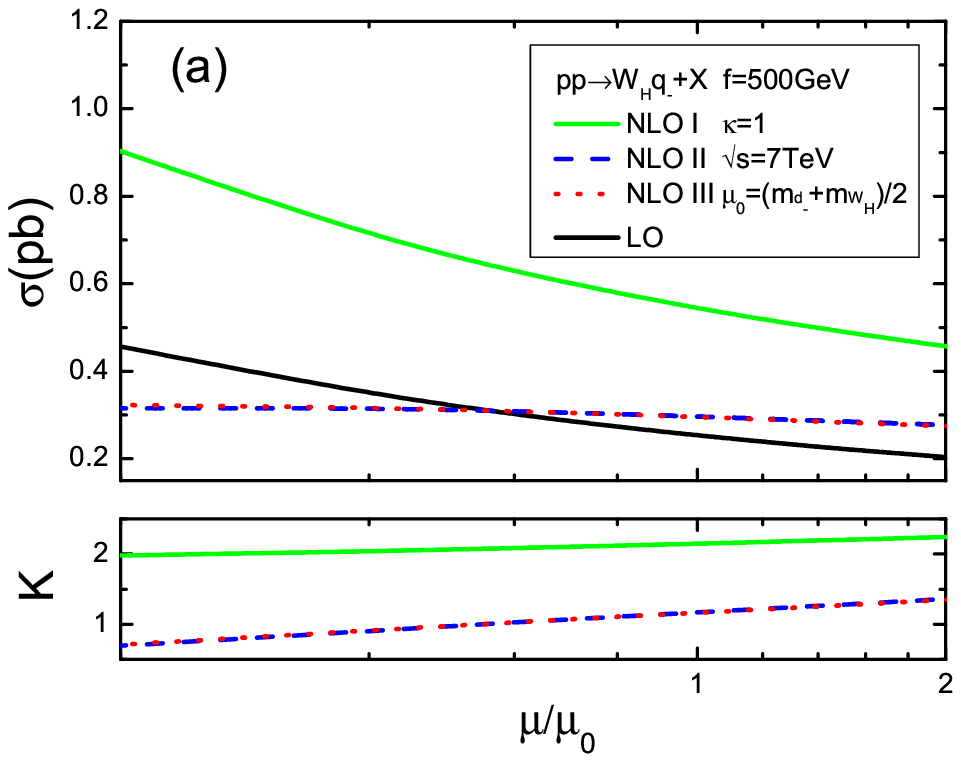}
\includegraphics[width=0.45\textwidth]{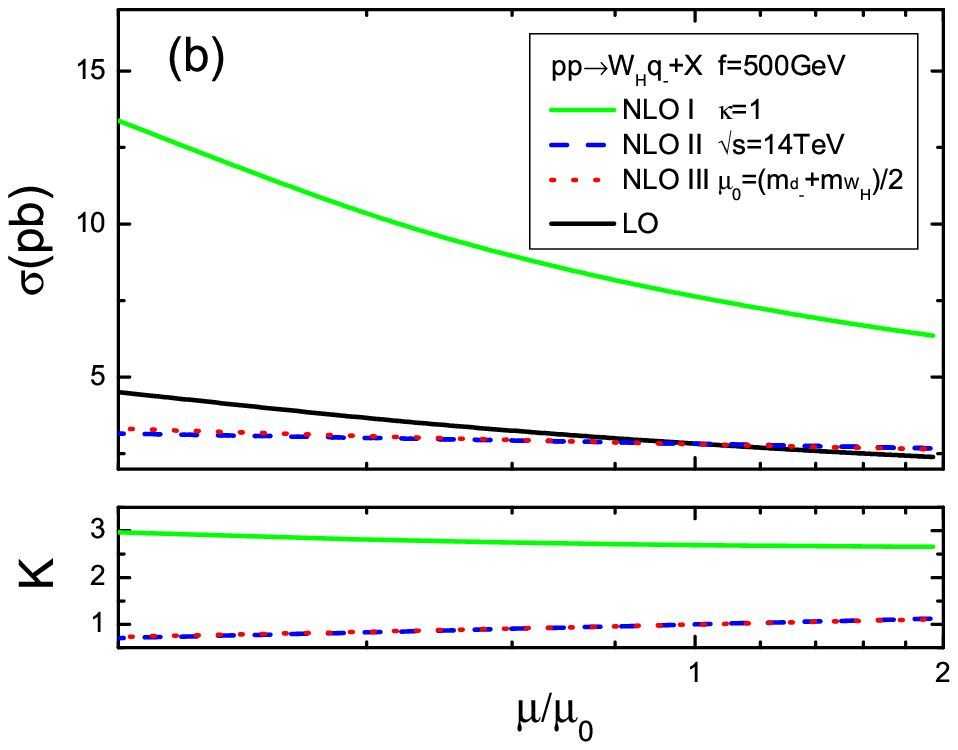}
\includegraphics[width=0.45\textwidth]{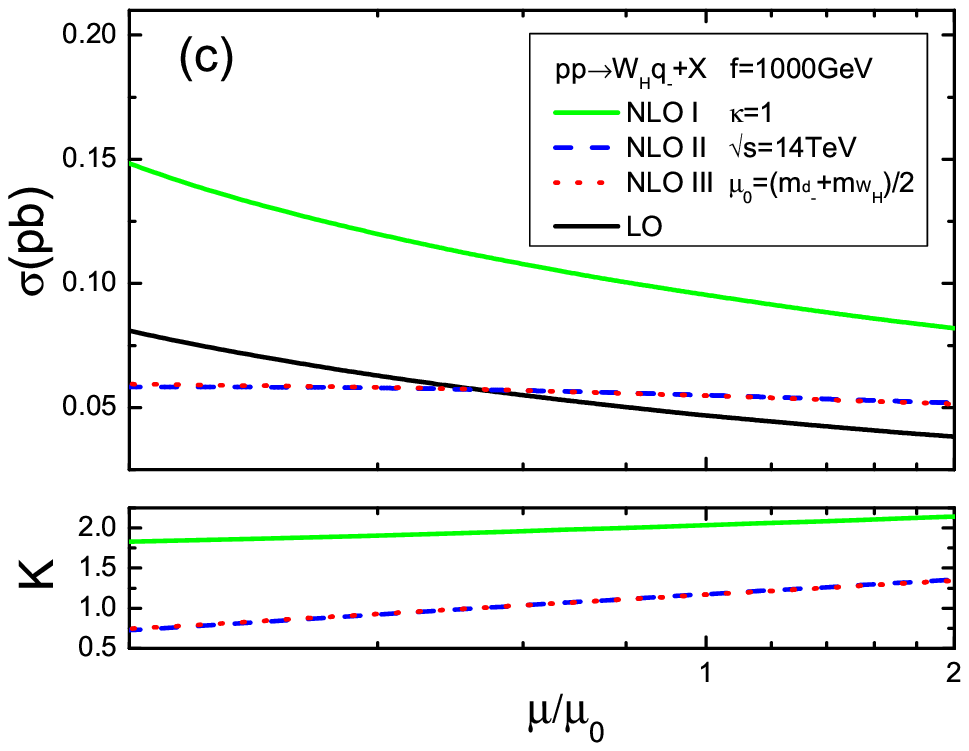}
\caption{\label{fig5} The dependence of the cross sections and the
corresponding K-factors for the \ppwq
process on the factorization/renormalization scale $\mu$ at the LHC.
(a) $f=500~GeV$, $\kappa=1$ and $\sqrt{s}=7~TeV$. (b) $f=500~GeV$,
$\kappa=1$ and $\sqrt{s}=14~TeV$. (c) $f=1~TeV$, $\kappa=1$ and
$\sqrt{s}=14~TeV$.  }
\end{center}
\end{figure}
\begin{figure}[htbp]
\begin{center}
\includegraphics[width=0.45\textwidth]{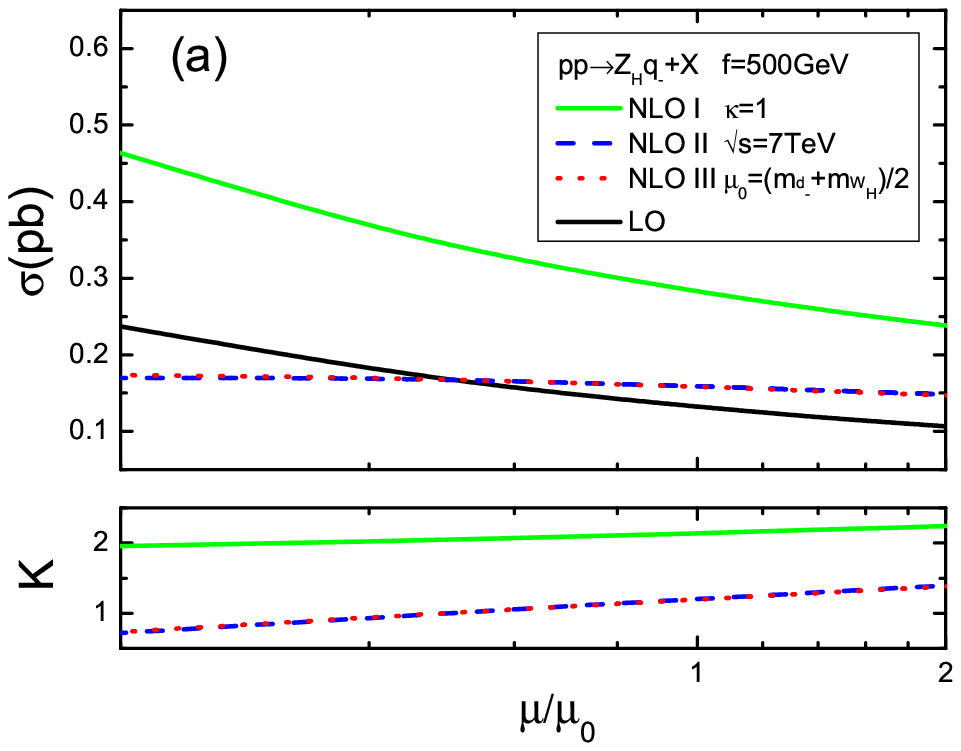}
\includegraphics[width=0.45\textwidth]{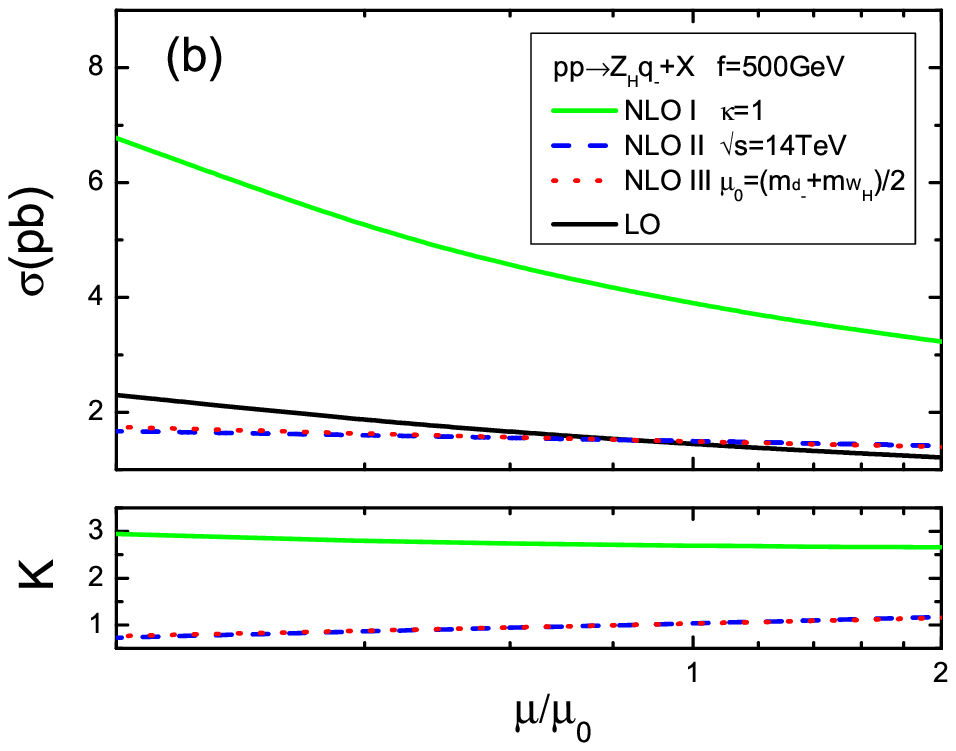}
\includegraphics[width=0.45\textwidth]{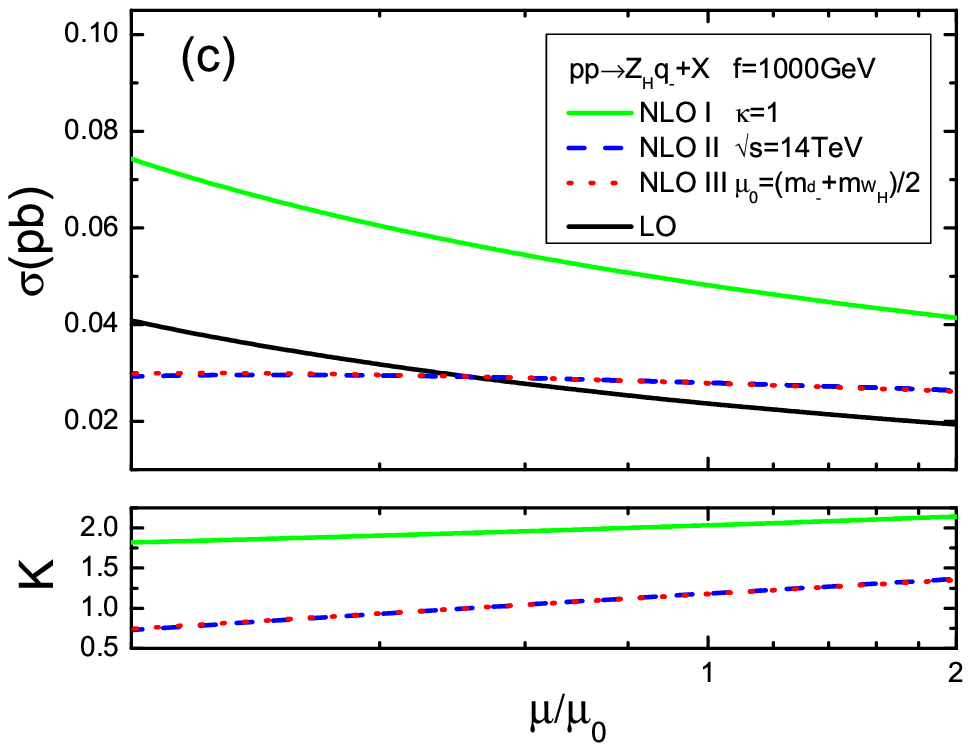}
\caption{\label{fig6} The dependence of the cross sections and the
corresponding K-factors for the \ppzq
process on the factorization/renormalization scale $\mu$ at the LHC.
(a) $f=500~GeV$, $\kappa=1$ and $\sqrt{s}=7~TeV$. (b) $f=500~GeV$,
$\kappa=1$ and $\sqrt{s}=14~TeV$. (c) $f=1~TeV$, $\kappa=1$ and
$\sqrt{s}=14~TeV$.  }
\end{center}
\end{figure}

In Figs.\ref{fig5}(a,b,c) and Figs.\ref{fig6}(a,b,c) we present the
LO, QCD NLO corrected cross sections and the corresponding K-factors
for the \ppwq and \ppzq processes as the functions of the
factorization/renormalization scale at the LHC with $\sqrt{s}=7~TeV$
and $14~TeV$, respectively. In Figs.\ref{fig5}(a,b) and
Figs.\ref{fig6}(a,b) the LHT input parameters are taken as
$f=500~GeV$ and $\kappa=1$, while in Fig.\ref{fig5}(c) and
Fig.\ref{fig6}(c) we take $f=1~TeV$ and $\kappa=1$. The masses of
$W_H$, $Z_H$ and $q_-$ ($q=u,d,c,s$) corresponding to these LHT
parameters are presented in Table \ref{table for mass parameters}.
In these figures the curves labeled by "NLO I", "NLO II" and "NLO
III" are for the QCD NLO corrected cross sections using the (I),
(II) and (III) schemes, respectively. The figures show that by using
the (II) and (III) subtraction schemes we can get almost the same
and moderate QCD NLO corrections to the production rate with a
strongly reduced factorization/renormalization scale uncertainty in
the plotted range of $\mu$, while the QCD NLO corrections using the
scheme (I) do not obviously improve the scale dependence of the LO
cross section and destroy the perturbative convergence in some range
of $\mu$. In the following analysis we set the
factorization/renormalization scale $\mu$ as its central value
$\mu_0=(m_{W_H}+m_{d_-})/2$.

\begin{table}
\begin{center}
\begin{tabular}{c|c|c|c|c|c}
  \hline
  $\kappa$ &  $f$    &  $m_{W_H}=m_{Z_H}$ & $m_{u_-}=m_{c_-}$  & $m_{d_-}=m_{s_-}$ & $\mu_0$ \\
  $~~~~~~~$  & $~~(GeV)~~$ & $~~(GeV)~~$  & $~~(GeV)~~$ & $~~(GeV)~~$ & $~~(GeV)~~$ \\
  \hline
    & 500  & 322.1  & 685.7   & 707.1   & 514.6   \\
    & 700  & 457.8  & 974.7   & 989.9   & 723.9   \\
  1 & 900  & 592.3  & 1260.9  & 1272.8  & 932.5  \\
    & 1000 & 659.3  & 1403.5  & 1414.2  & 1036.7 \\
    & 1100 & 726.1  & 1545.9  & 1555.6  & 1140.9 \\
    & 1300 & 859.7  & 1830.3  & 1838.5  & 1349.1 \\
  \hline
   & 500 & 322.1 & 2057.1   & 2121.3   & 1221.7 \\
 3 & 700 & 457.8 & 2924.0   & 2969.9   & 1713.8 \\
   & 900 & 592.3 & 3782.7   & 3818.4   & 2205.3 \\
   \hline
\end{tabular}
\end{center}
\begin{center}
\begin{minipage}{15cm}
\caption{\label{table for mass parameters} The masses of $W_H$, $Z_H$ and
$q_-$ ($q=u,d,c,s$) for some typical values of the LHT parameters.  }
\end{minipage}
\end{center}
\end{table}

\par
\subsection{Dependence on LHT parameters \label{sub-1} }
\par
\begin{figure}[htbp]
\begin{center}
\includegraphics[width=0.45\textwidth]{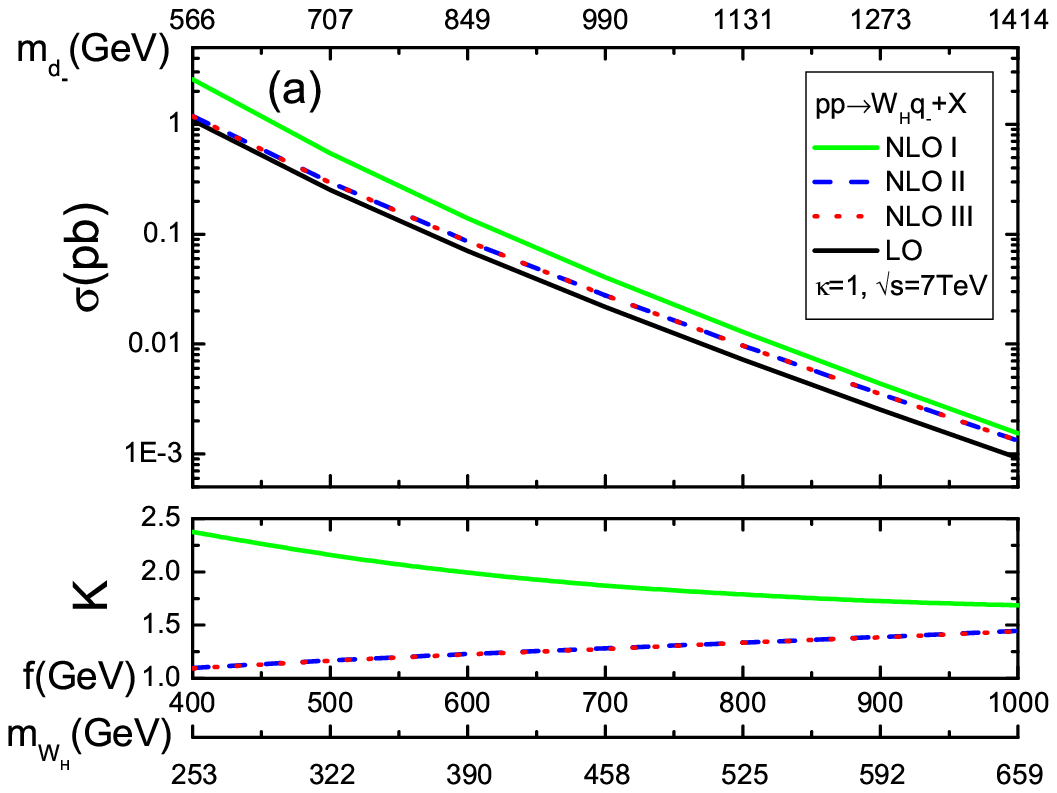}
\includegraphics[width=0.45\textwidth]{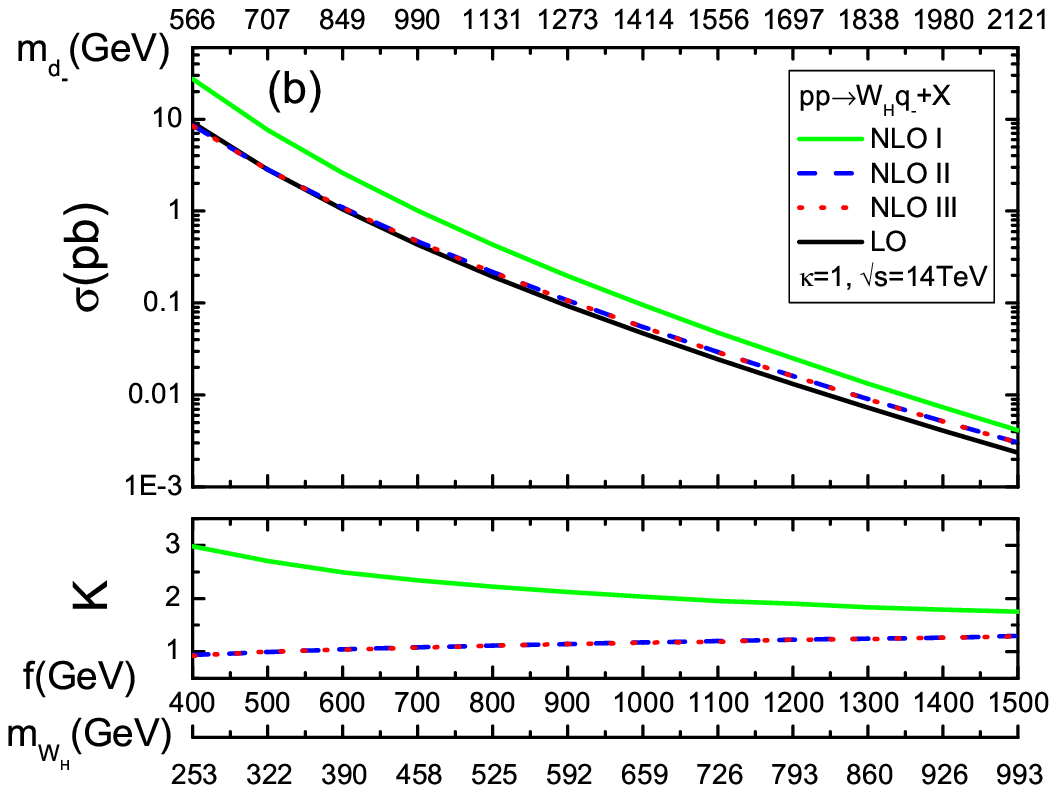}
\includegraphics[width=0.45\textwidth]{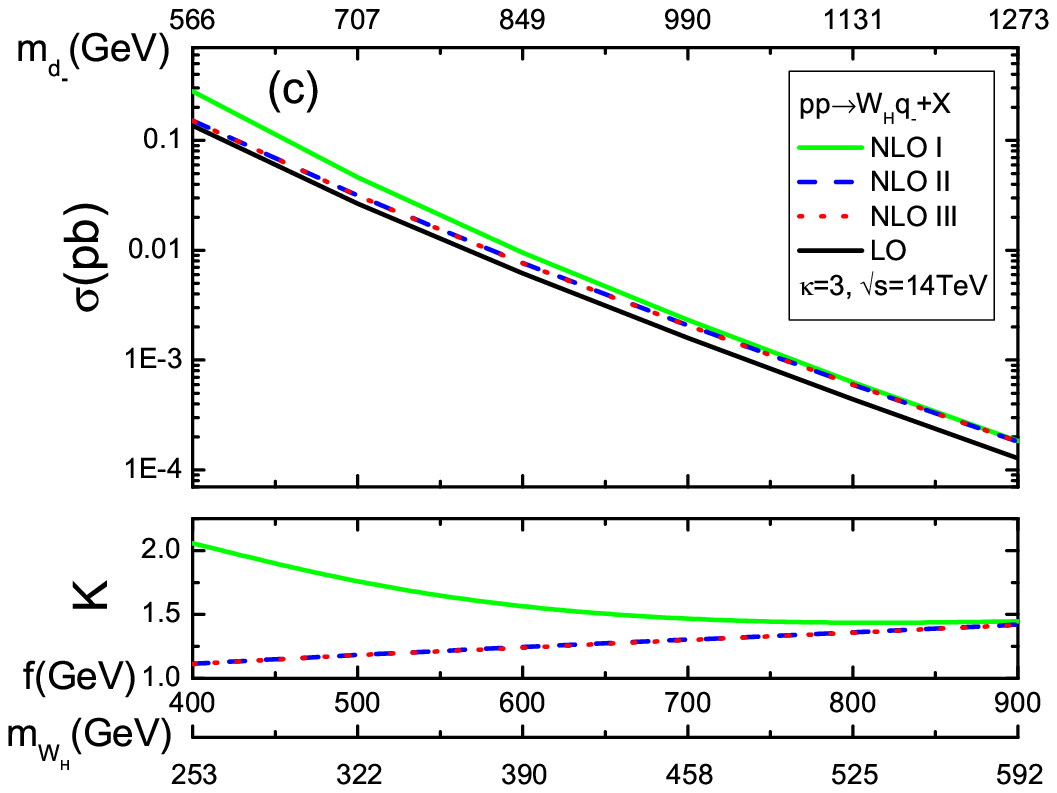}
\caption{\label{fig7} The cross sections and the corresponding
K-factors for the \ppwq process as the functions of the LHT
parameter $f$ at the LHC. The corresponding $m_{W_H}$ and $m_{d_-}$
values are also scaled on the x-axis. (a) $\kappa=1$ and
$\sqrt{s}=7~TeV$. (b) $\kappa=1$ and $\sqrt{s}=14~TeV$. (c)
$\kappa=3$ and $\sqrt{s}=14~TeV$. }
\end{center}
\end{figure}
\begin{figure}[htbp]
\begin{center}
\includegraphics[width=0.45\textwidth]{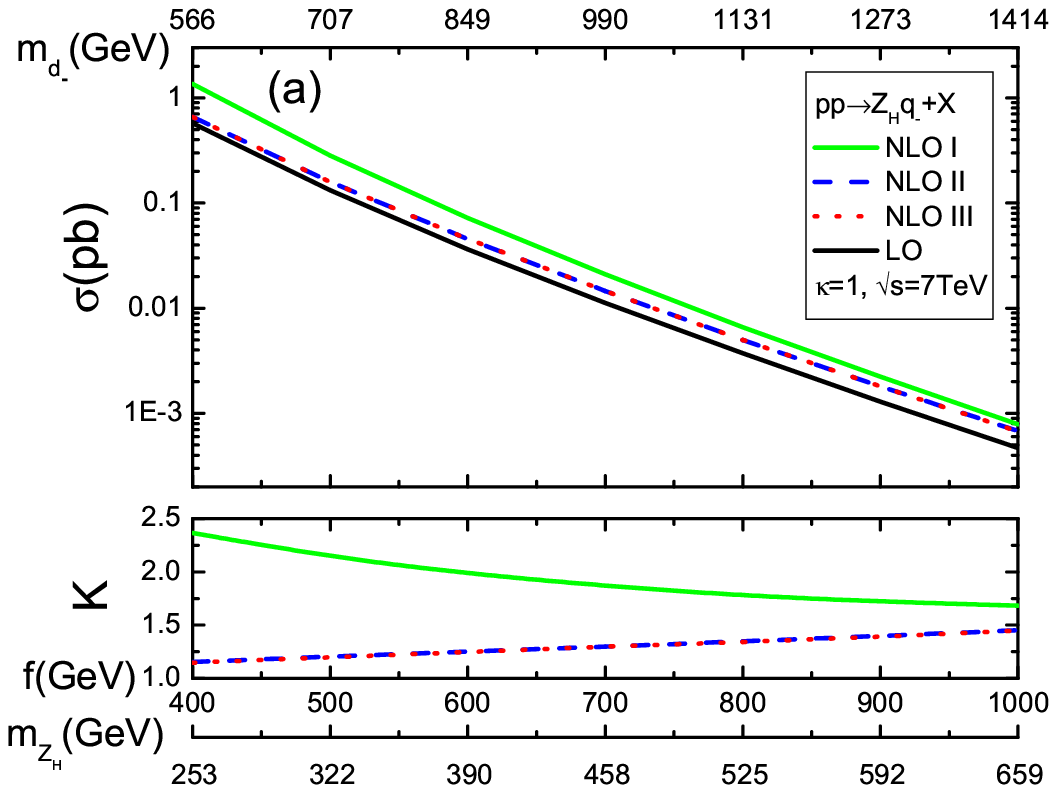}
\includegraphics[width=0.45\textwidth]{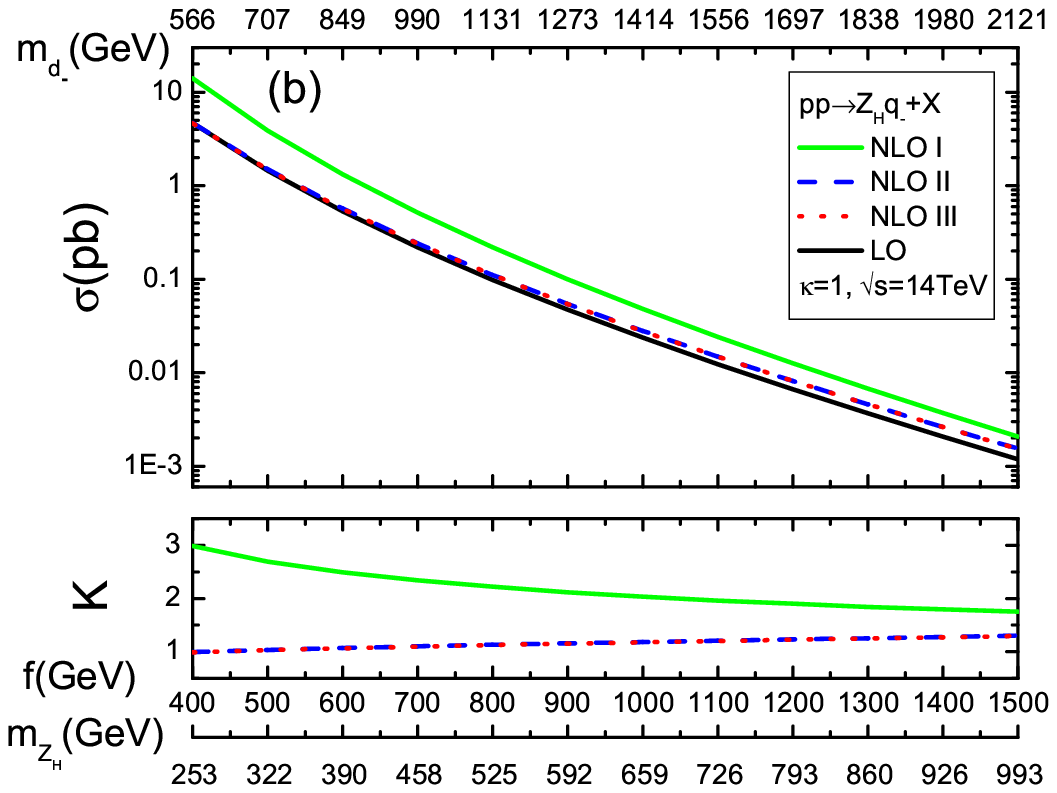}
\includegraphics[width=0.45\textwidth]{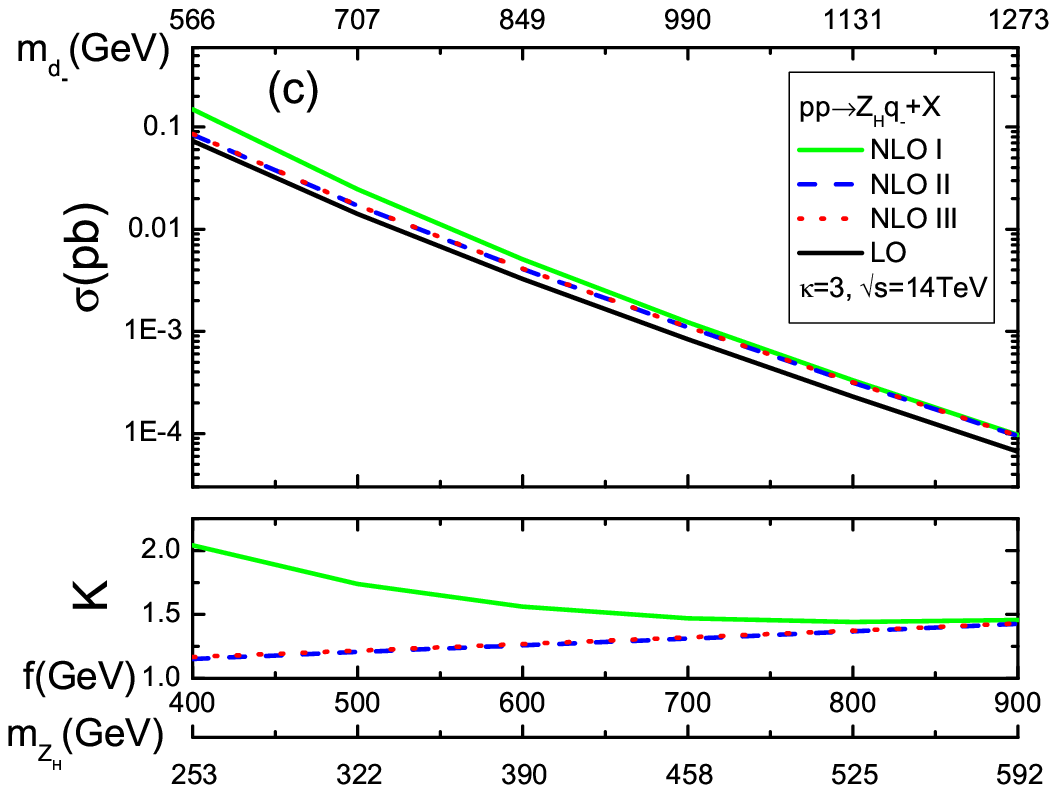}
\caption{\label{fig8} The cross sections and the corresponding
K-factors for the \ppzq process as the functions of the LHT
parameter $f$ at the LHC. The corresponding $m_{W_H}$ and $m_{d_-}$
values are also scaled on the x-axis. (a) $\kappa=1$ and
$\sqrt{s}=7~TeV$. (b) $\kappa=1$ and $\sqrt{s}=14~TeV$. (c)
$\kappa=3$ and $\sqrt{s}=14~TeV$. }
\end{center}
\end{figure}

We depict the LO, QCD NLO corrected cross sections and the
corresponding K-factors for the \ppwq and \ppzq processes as the
functions of $f$, the $SU(5)$ global symmetry breaking scale of the
LHT, at the LHC with $\sqrt{s}=7~TeV$ and $14~TeV$ in
Figs.\ref{fig7}(a,b,c) and Figs.\ref{fig8}(a,b,c), respectively. In
Figs.\ref{fig7}(a,b) and Figs.\ref{fig8}(a,b) the parameter $\kappa$
is set to be $1$, while in Fig.\ref{fig7}(c) and Fig.\ref{fig8}(c)
we take $\kappa=3$. The curves labeled by "NLO I", "NLO II" and "NLO
III" are for the QCD NLO corrected cross sections using the (I),
(II) and (III) schemes, respectively. One can conclude from these
figures that the cross section for the $pp \rightarrow W_H(Z_H) q_-
+ X$ process decreases quickly with the increment of $f$, because
the two final T-odd particles become heavier with the increment of
$f$. However, in the plotted range of $f$ we could have observable
production rates for the $p p \to W_H q_- + X$ and $p p \to Z_H q_-
+ X$ processes, especially when $\kappa=1$.

\par
We present the numerical results for the $pp \to W_H q_- + X$ and
$pp \to Z_H q_- + X$ processes at the LHC for some typical values of
the LHT parameters in Table \ref{tab2} and Table \ref{tab3},
respectively. In the two tables we list the values of the
proton-proton colliding energy $\sqrt{s}$, the LHT parameters
$\kappa$ and $f$, the LO cross section $\sigma_{LO}$, the QCD NLO
corrected cross sections using the (I), (II) and (III) schemes, and
the corresponding K-factors. From both the tables and the
$f$-dependence figures, we can see that the QCD NLO corrected cross
sections for the $pp \rightarrow W_H q_- + X$ and $pp \rightarrow
Z_H q_- + X$ processes at the LHC by using the subtraction scheme
(II) are almost the same as those by adopting the subtraction scheme
(III) in the LHT parameter space considered in this paper. At the
early LHC, the QCD NLO corrected cross section for the $pp
\rightarrow W_H q_- + X$ ($pp \rightarrow Z_H q_- + X$) process
using the subtraction scheme (III) can reach $295.3fb$ ($158.2fb$)
and the corresponding K-factor is $1.16$ ($1.20$) when $f =
500~{GeV}$ and $\kappa=1$. While at the $14~TeV$ LHC, the QCD NLO
corrected cross section for the $pp \rightarrow W_H q_- + X$ ($pp
\rightarrow Z_H q_- + X$) process using the subtraction scheme (III)
can reach $2832fb$ ($1491fb$) and the corresponding K-factor is
$1.00$ ($1.03$) when $f = 500~{GeV}$ and $\kappa=1$. We can also
find that the QCD NLO K-factor increases with the increment of the
LHT parameters $\kappa$ and $f$. By adopting the subtraction scheme
(III), the K-factor can reach $1.42$ and $1.43$ when $\kappa = 3$
and $f = 900~GeV$ for the $pp \rightarrow W_H q_- + X$ and $pp
\rightarrow Z_H q_- + X$ processes at the $14~TeV$ LHC,
respectively.

\begin{table}
\begin{center}
\begin{tabular}{c|c|c|c|c|c|c|c|c|c}
  \hline
  $\sqrt{s}$  &  $\kappa$  & $f$ & $\sigma_{LO}$  & $\sigma_{NLO}^{(I)}$ & $K^{(I)}$
  & $\sigma_{NLO}^{(II)}$ & $K^{(II)}$ & $\sigma_{NLO}^{(III)}$ & $K^{(III)}$ \\
  $(TeV)$ & ~~~~ & $(GeV)$ & $(fb)$ & $(fb)$ & &$(fb)$ & & $(fb)$ & \\
  \hline
    &  & 500 & 253.53(1)     &544.6(8)  &2.15   &296.6(8) &1.17   &295.3(8)  &1.16 \\
  7 &1 & 700 & 21.721(1)     &40.52(7)  &1.87   &27.87(7) &1.28   &27.73(7)  &1.28 \\
    &  & 900 & 2.5287(1)     &4.352(9)  &1.72   &3.514(9) &1.39   &3.503(9)  &1.39 \\
\hline
    &  & 500  & 2830.7(1)    &7648(3)       &2.70   &2823(2)   &1.00   &2832(3)   &1.00 \\
    &  & 700  & 432.10(2)    &1011.9(4)     &2.34   &467.9(3)  &1.08   &465.8(3)  &1.08 \\
 14 &1 & 900  & 93.171(3)    &197.66(8)     &2.12   &106.86(7) &1.14   &106.45(8) &1.14 \\
    &  & 1100 & 24.460(1)    &47.82(2)      &1.96   &29.32(2)  &1.20   &29.08(2)  &1.19 \\
    &  & 1300 & 7.2751(3)    &13.349(6)     &1.83   &9.053(6)  &1.24   &8.995(7)  &1.24 \\
  \hline
    &  & 500 & 26.543(1)     &46.14(6)   &1.74   &31.42(5)     &1.18   &31.42(6)  &1.18 \\
 14 &3 & 700 & 1.5903(1)     &2.320(6)   &1.46   &2.072(6)     &1.30   &2.067(6)  &1.30 \\
    &  & 900 & 0.12741(1)    &0.1841(5)  &1.44   &0.1810(5)    &1.42   &0.1806(5) &1.42 \\
   \hline
\end{tabular}
\end{center}
\begin{center}
\begin{minipage}{15cm}
\caption{\label{tab2} The numerical results for the $pp \to W_H q_-
+ X$ process at the LHC for some typical values of the LHT
parameters. }
\end{minipage}
\end{center}
\end{table}
\begin{table}
\begin{center}
\begin{tabular}{c|c|c|c|c|c|c|c|c|c}
  \hline
  $\sqrt{s}$  &  $\kappa$  & $f$ & $\sigma_{LO}$  & $\sigma_{NLO}^{(I)}$ & $K^{(I)}$
  & $\sigma_{NLO}^{(II)}$ & $K^{(II)}$ & $\sigma_{NLO}^{(III)}$ & $K^{(III)}$ \\
  $(TeV)$ & ~~~~ & $(GeV)$ & $(fb)$ & $(fb)$ & &$(fb)$ & & $(fb)$ & \\
  \hline
    &  & 500   & 132.15(1)     & 282.7(6)  &2.14   &158.9(6)  &1.20  &158.2(7)  &1.20 \\
  7 &1 & 700   & 11.205(1)     & 20.94(5)  &1.87   &14.54(5)  &1.30  &14.54(6)  &1.30 \\
    &  & 900   & 1.2977(1)     & 2.232(6)  &1.72   &1.813(6)  &1.40  &1.808(6)  &1.39 \\
\hline
    &  & 500   & 1446.5(1)     &3898(6)    &2.69   &1497(6)   &1.03  &1491(6)   &1.03 \\
    &  & 700   & 219.36(2)     &513.6(9)   &2.34   &241.6(9)  &1.10  &240.6(9)  &1.10 \\
  14 &1 & 900  & 47.141(4)     &99.9(2)    &2.12   &54.6(2)   &1.16  &54.3(2)   &1.15 \\
    &  & 1100  & 12.351(1)     &24.22(6)   &1.96   &14.89(6)  &1.21  &14.83(6)  &1.20 \\
    &  & 1300  & 3.6685(3)     &6.758(9)   &1.84   &4.598(9) &1.25  &4.582(9)   &1.25 \\
  \hline
    &  & 500  & 14.125(1)     &24.54(7)    &1.74   &17.02(6)    &1.21 &17.18(7)    &1.22 \\
  14 &3 & 700 & 0.83522(7)    &1.227(4)    &1.47   &1.095(5)    &1.31 &1.102(4)    &1.32 \\
    &  & 900  & 0.066474(5)   &0.0969(3)   &1.46   &0.0947(3)   &1.42 &0.0952(3)   &1.43 \\
   \hline
\end{tabular}
\end{center}
\begin{center}
\begin{minipage}{15cm}
\caption{\label{tab3} The numerical results for the $pp \to Z_H q_-
+ X$ process at the LHC for some typical values of the LHT
parameters. }
\end{minipage}
\end{center}
\end{table}

\vskip 5mm
\subsection{Transverse momentum and rapidity distributions of final particles }
\par
In this subsection we inspect the characteristics of the transverse
momentum and rapidity distributions of the final decay products. The
$W_Hq_-$ associated production at the LHC can be followed by the
subsequent decays $W_H \to A_H W$, $q_- \to W_H q' \to A_H W q'$ and
$W^{\mp} \to l^{\mp}\stackrel{(-)}{\nu}$. Therefore, the $W_Hq_-$
production signal can be found by detecting the final states $l^+
l^- + {\rm jet} + E_{T,{\rm missing}}$ ($E_{T,{\rm missing}} = A_H
A_H \nu \bar{\nu}$). Similarly, the $Z_H q_-$ production can be
detected through the decays $Z_H \to A_H H$, $H \to b\bar b$ and
$q_- \to W_H q' \to A_H W q' \to A_H l \nu q'$, with the final
states as $l^{\mp} b \bar{b} + {\rm jet} + E_{T,{\rm missing}}$
($E_{T,{\rm missing}} =A_HA_H \stackrel{(-)}{\nu}$).

\par
The results in subsections  \ref{sub-0} and \ref{sub-1} show that
the QCD NLO corrections using the scheme (I) would destroy the
convergence of the perturbative QCD description of the $W_H(Z_H)q_-$
associated production, while the QCD corrections using the (II) and
(III) subtraction schemes are almost the same and can keep the
perturbative convergence. In the following discussions on the
transverse momentum ($p_T$) and rapidity ($y$) distributions of
final particles, we present only the numerical results by adopting
the subtraction scheme (III).

\par
\begin{figure}[htbp]
\begin{center}
\includegraphics[width=0.45\textwidth]{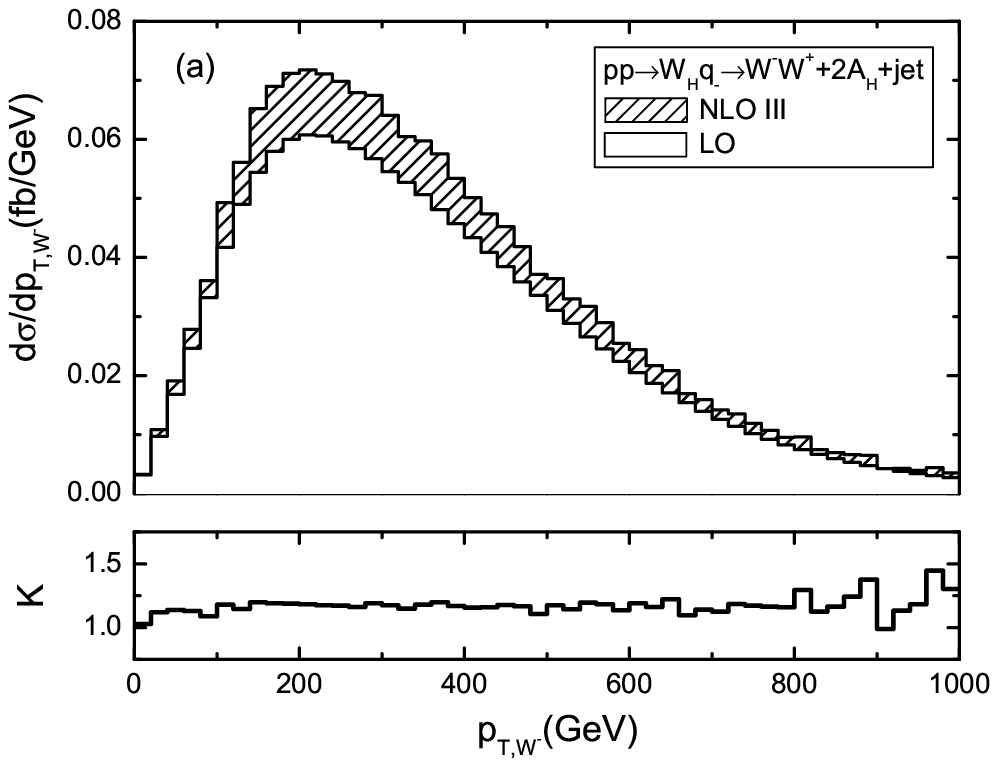}
\includegraphics[width=0.45\textwidth]{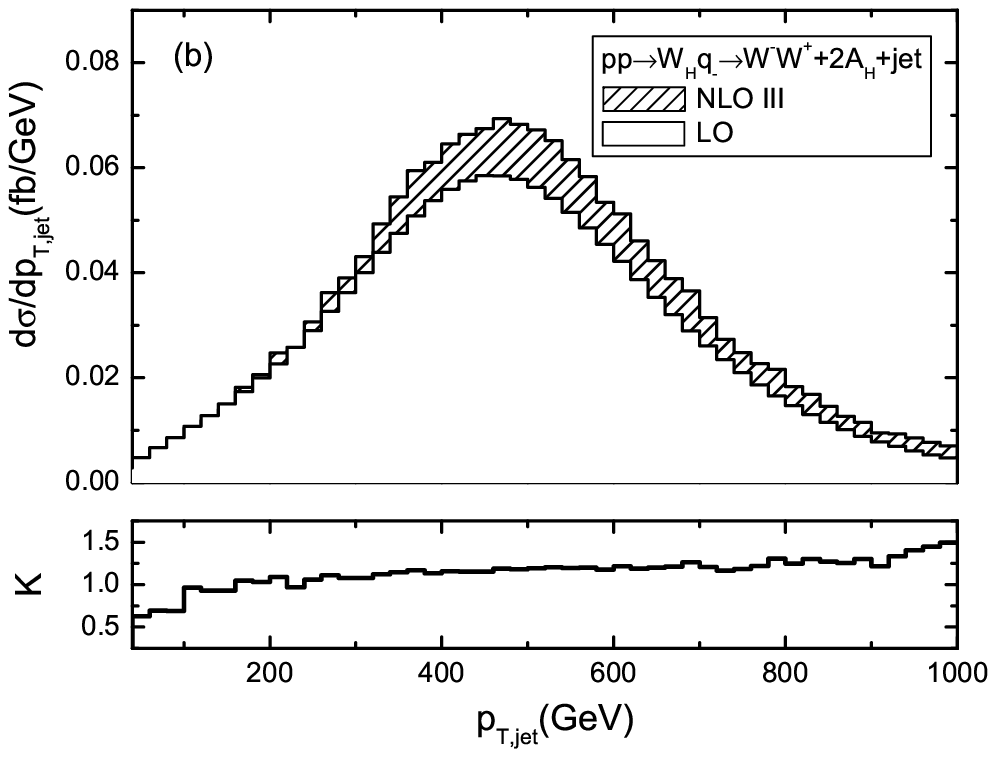}
\includegraphics[width=0.45\textwidth]{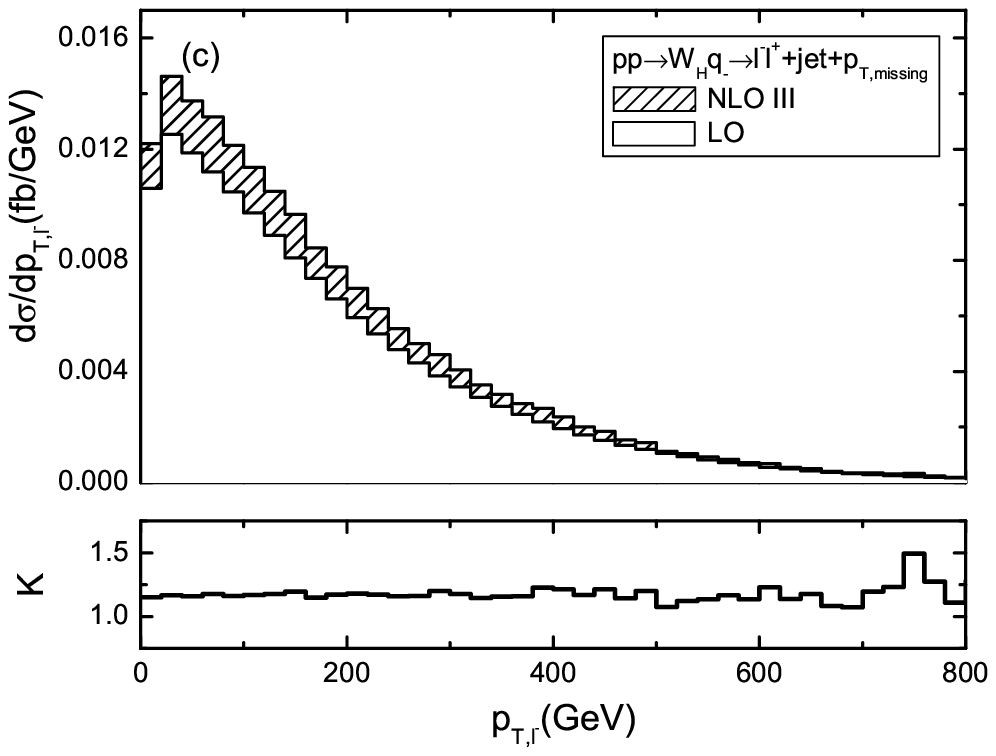}
\includegraphics[width=0.45\textwidth]{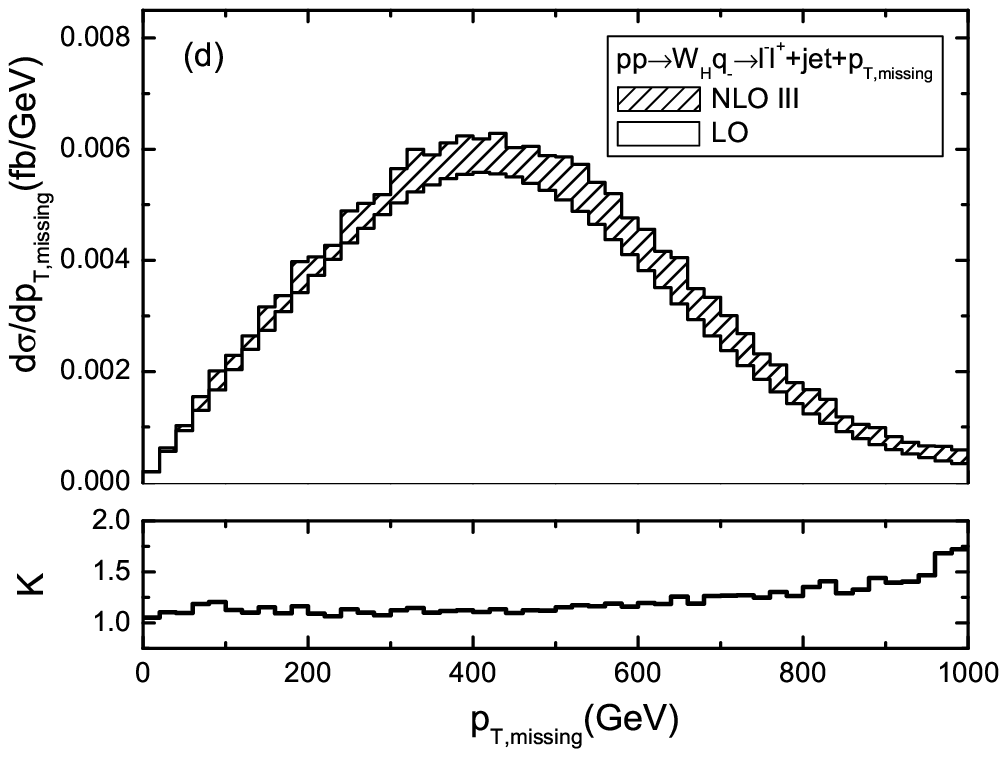}
\caption{\label{fig9} The LO and QCD NLO corrected $p_T$
distributions of final particles for the \ppwq process at the LHC by
taking $f=1~TeV$, $\kappa=1$ and $\sqrt{s}=14~TeV$.
(a) $W^{-}$-boson, (b) jet, (c) lepton $l^-$, (d) missing $p_T$. }
\end{center}
\end{figure}

The LO and QCD NLO corrected transverse momentum distributions of
the final $W^-$-boson, jet, lepton $l^-$ and missing energy ($A_HA_H
\nu \bar{\nu}$) for the $pp \rightarrow W_H q_- + X$ process and the
corresponding K-factors are plotted in Figs.\ref{fig9}(a,b,c,d),
respectively. The results in these figures are obtained by taking
$f=1~TeV$, $\kappa=1$ and $\sqrt{s}=14~TeV$. With these parameters
we obtain $m_{A_H} = 153.0~GeV$ from Eq.(\ref{mass-AH-VH}). It
should be declared that in Fig.\ref{fig9}(b) the $p_T$ distribution
labeled by "NLO III" is for the leading jet ($j_1$ is called as
leading jet if $E_{j_1}>E_{j_2}$.), if there exist two jets in one
event. From these figures we find that the distributions $\frac{d
\sigma}{d p_{T, W^-}}$, $\frac{d \sigma}{d p_{T, {\rm jet}}}$ and
$\frac{d \sigma}{d p_{T, {\rm missing}}}$ increase with the
increment of $p_T$ in the low $p_T$ region, and reach their maxima
at $p_{T, W^-} \sim 200~GeV$, $p_{T, {\rm jet}} \sim 500~GeV$ and
$p_{T, {\rm missing}} \sim 400~GeV$, respectively. The transverse
momentum distribution of $l^-$ is quite different from the former
ones. It decreases rapidly with the increment of $p_{T, l^-}$ for
experimentally acceptable lepton.

\par
\begin{figure}[htbp]
\begin{center}
\includegraphics[width=0.45\textwidth]{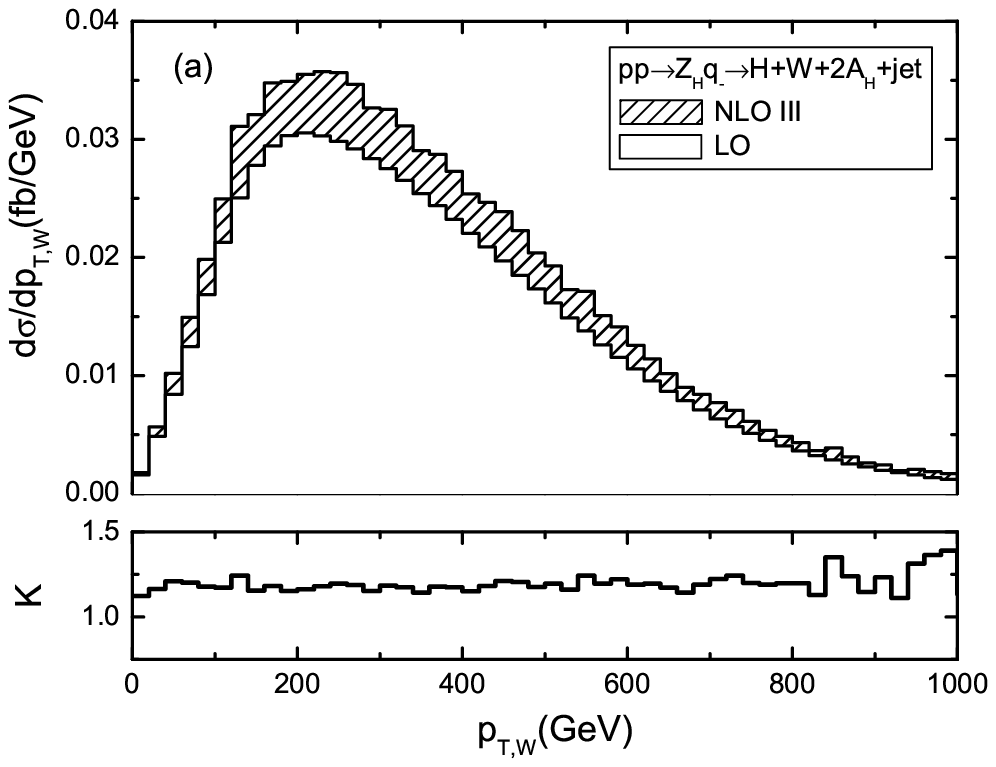}
\includegraphics[width=0.45\textwidth]{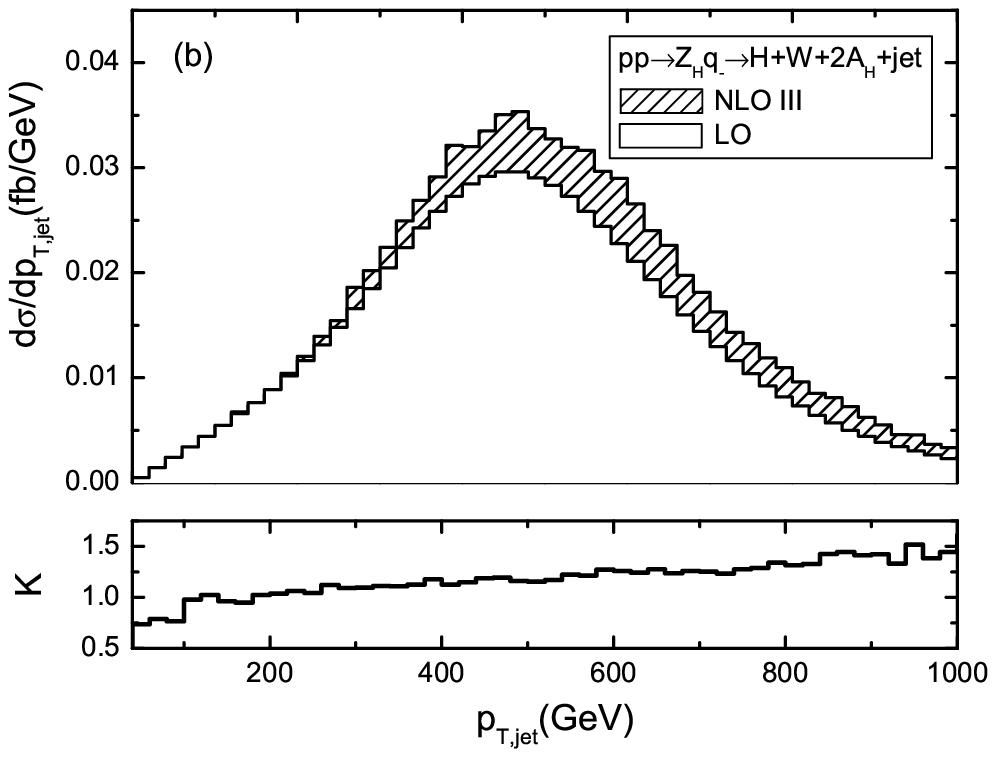}
\includegraphics[width=0.45\textwidth]{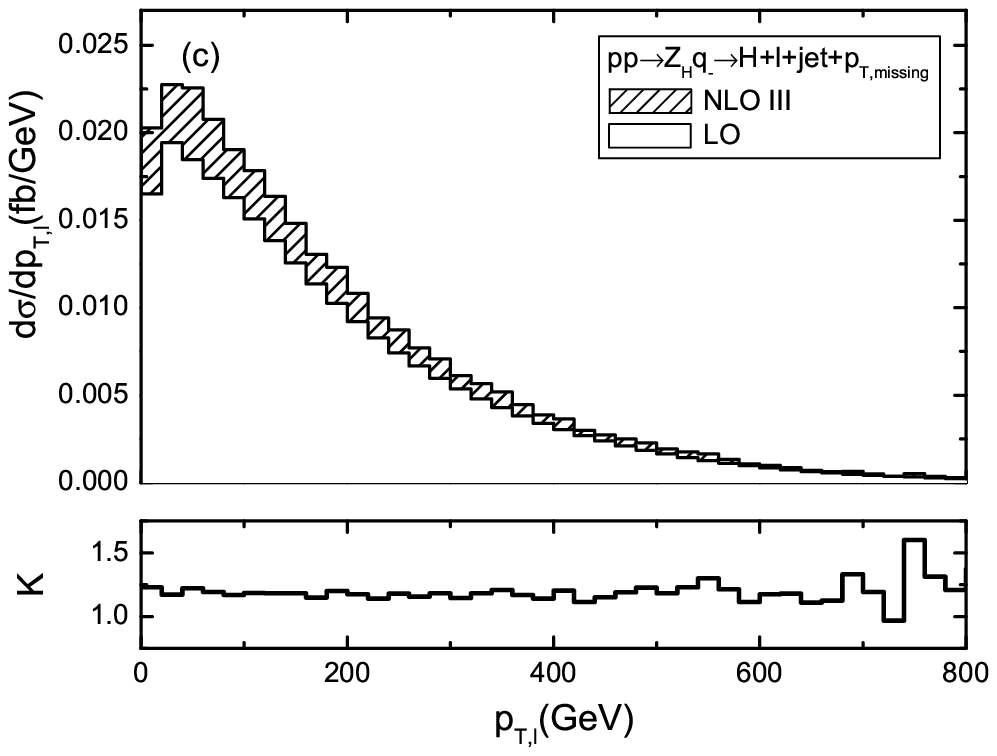}
\includegraphics[width=0.45\textwidth]{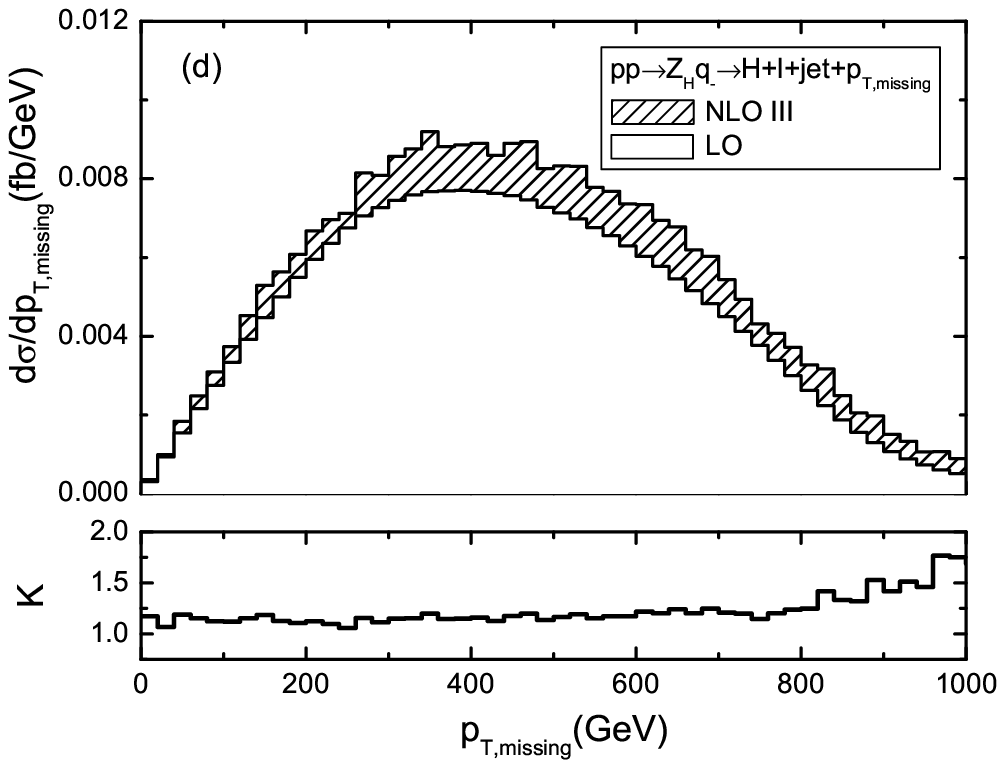}
\caption{\label{fig10} The LO and QCD NLO corrected $p_T$
distributions of final particles for the \ppzq process at the LHC by
taking $f=1~TeV$, $\kappa=1$
and $\sqrt{s}=14~TeV$. (a) $W$-boson, (b) jet, (c) lepton, (d) missing $p_T$. }
\end{center}
\end{figure}

The LO and QCD NLO corrected transverse momentum distributions of
the final $W$-boson, jet, lepton and missing energy ($A_HA_H
\stackrel{(-)}{\nu}$) for the $pp \rightarrow Z_H q_- + X$ process
and the corresponding K-factors are depicted in
Figs.\ref{fig10}(a,b,c,d), respectively. We do not distinguish the
electric charge of $W^{\pm}$ and $l^{\pm}$ in the $p_T$
distributions of the final $W$-boson and lepton. In these figures we
set the input parameters as $f=1~TeV$, $\kappa=1$ and
$\sqrt{s}=14~TeV$. Again the $p_T$ distribution labeled by "NLO III"
in Fig.\ref{fig10}(b) is for the leading jet in a two-jet event.
These figures show that the transverse momentum distributions of
$W^{\pm}$, jet, $l^{\pm}$ and missing energy for the $pp \rightarrow
Z_H q_- + X$ process are quite similar to those of $W^-$, jet, $l^-$
and missing energy for the $pp \rightarrow W_H q_- + X$ process,
respectively, while the production rate of the $pp \rightarrow W_H
q_- + X$ process is almost twice larger than that of the $pp
\rightarrow Z_H q_- + X$ process.

\par
\begin{figure}[htbp]
\begin{center}
\includegraphics[width=0.45\textwidth]{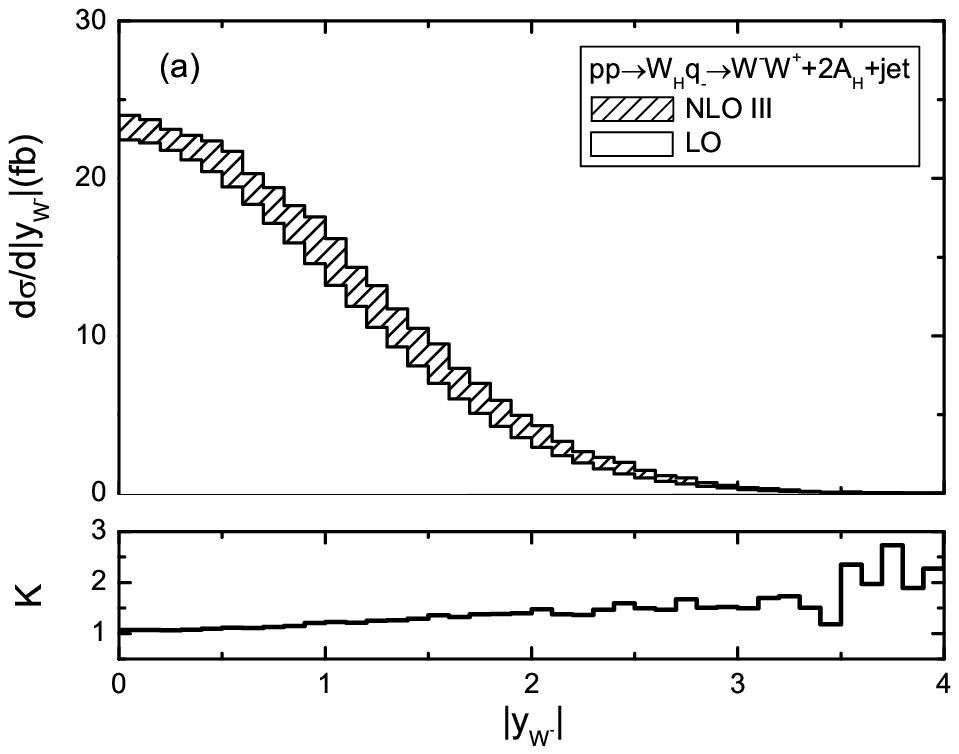}
\includegraphics[width=0.45\textwidth]{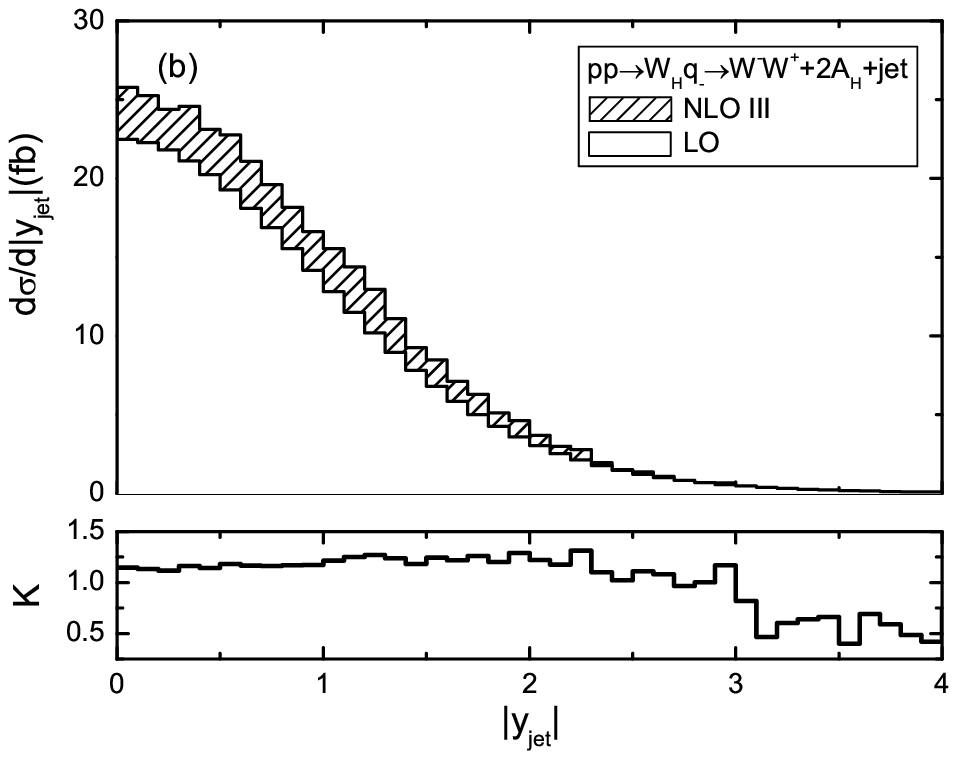}
\includegraphics[width=0.45\textwidth]{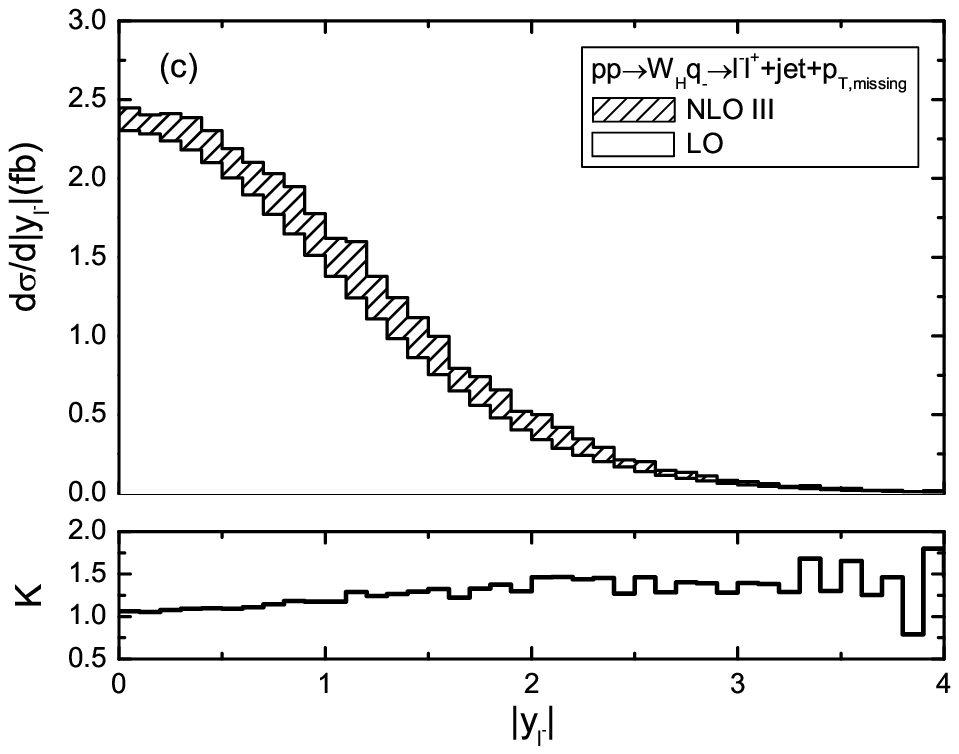}
\caption{\label{fig11} The LO and QCD NLO corrected rapidity
distributions of final particles for the \ppwq process at the LHC by
taking $f=1~TeV$, $\kappa=1$
and $\sqrt{s}=14~TeV$. (a) $W^{-}$-boson, (b) jet, (c) lepton $l^-$. }
\end{center}
\end{figure}
\begin{figure}[htbp]
\begin{center}
\includegraphics[width=0.45\textwidth]{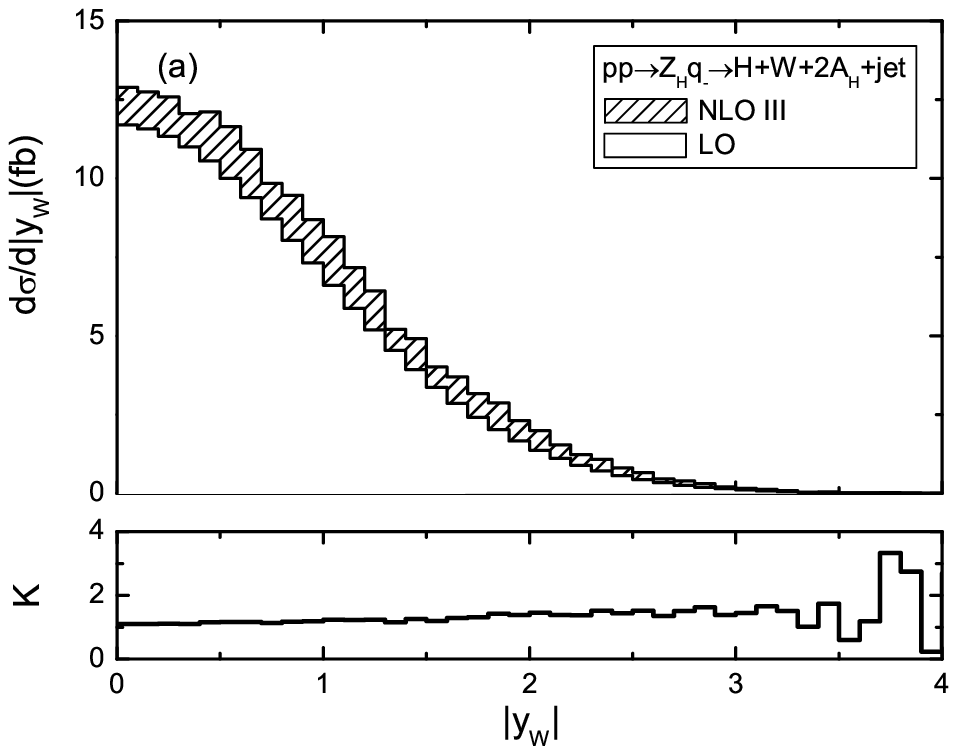}
\includegraphics[width=0.45\textwidth]{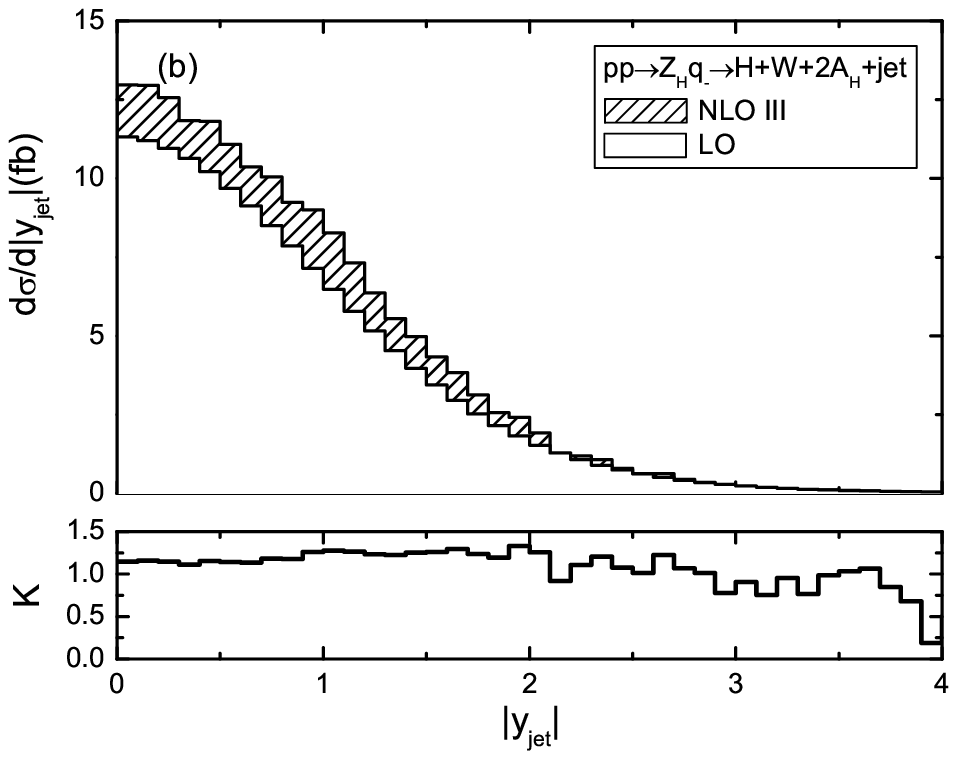}
\includegraphics[width=0.45\textwidth]{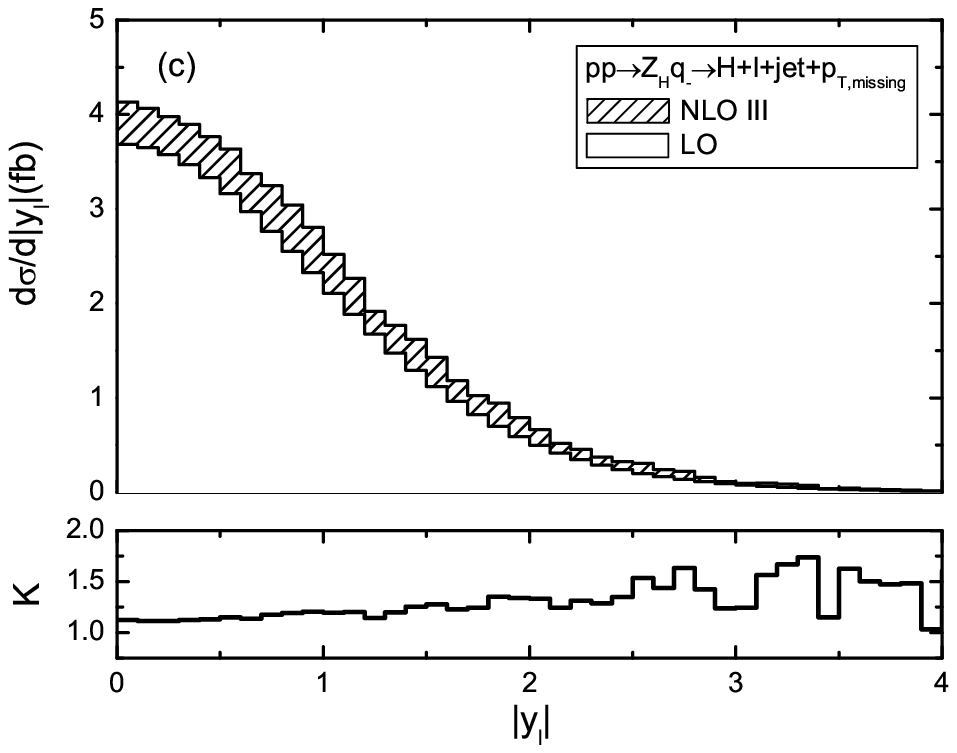}
\caption{\label{fig12} The LO and QCD NLO corrected rapidity
distributions of final particles for the \ppzq process at the LHC by
taking $f=1~TeV$, $\kappa=1$
and $\sqrt{s}=14~TeV$. (a) $W$-boson, (b) jet, (c) lepton. }
\end{center}
\end{figure}

We present the LO and QCD NLO corrected rapidity distributions of
final particles and the corresponding K-factors for the $pp
\rightarrow W_H q_- + X$ and $pp \rightarrow W_H q_- + X$ processes
in Figs.\ref{fig11}(a,b,c) and Figs.\ref{fig12}(a,b,c),
respectively. Because $\frac{d \sigma}{dy}\Big|_{-y} = \frac{d
\sigma}{dy}\Big|_y$ for the processes at the LHC, we plot only the
distributions $\frac{d \sigma}{d|y|}$. The values of the input
parameters are taken the same as those used in Figs.\ref{fig9} and
Figs.\ref{fig10}. For the $Z_H q_-$ production, we do not
distinguish the electric charge of $W^{\pm}$ and $l^{\pm}$ in the
rapidity distributions of the final $W$-boson and lepton. We can see
from these figures that the rapidity distributions $\frac{d
\sigma}{d |y_{W^-}|}$, $\frac{d \sigma}{d |y_{{\rm jet}}|}$ and
$\frac{d \sigma}{d |y_{l^-}|}$ for the $pp \rightarrow W_H q_- + X$
process are quite similar to the distributions $\frac{d \sigma}{d
|y_{W}|}$, $\frac{d \sigma}{d |y_{{\rm jet}}|}$ and $\frac{d
\sigma}{d |y_{l}|}$ for the $pp \rightarrow Z_H q_- + X$ process,
respectively. All these differential cross sections decrease with
the increment of $|y|$. That means all the final products, including
$W$-boson, jet and lepton, prefer producing transversely.

\vskip 5mm
\section{Summary}
\par
In this paper, we calculate the $W_H(Z_H) q_-$
($q_-=u_-,d_-,c_-,s_-, \bar{u}_-, \bar{d}_-, \bar{c}_-,\bar{s}_-$)
associated production at the CERN LHC up to the QCD NLO. We
investigate theoretically the dependence of the cross section on the
factorization/renormalization scale, and present the transverse
momentum and rapidity distributions of the final decay products.

\par
The real light-quark emission partonic processes $gg \rightarrow
W_H(Z_H) q_-^{\prime} + \bar{q}$ and $q^{\prime \prime}
\bar{q}^{\prime \prime} \rightarrow W_H(Z_H) q_-^{\prime} + \bar{q}$
are at the same order of the on-shell T-odd quark pair production
with the subsequent decay $q_- \rightarrow W_H(Z_H) + q$. Including
these partonic processes will give a large contribution to the NLO
QCD corrected rate for the associated $W_H(Z_H)q_-$ production. In
order to show how to provide a reliable production rate of the $pp
\rightarrow W_H(Z_H) q_- + X$ process at the LHC, we adopt three
schemes in considering the QCD NLO corrections for comparison. Our
calculations demonstrate that by using the scheme (I) the
perturbative convergence could be destroyed due to the double
counting showing up, while we can keep the convergence of the
perturbative QCD description and obtain moderate QCD NLO corrections
to the production rate with a strongly reduced scale uncertainty by
adopting the (II) and (III) subtraction schemes. The smallness of
the discrepancy between the numerical results using the (II) and
(III) subtraction schemes indicates that the contributions of the
partonic processes $gg \rightarrow W_H(Z_H) q_-^{\prime} + \bar{q}$
and $q^{\prime \prime} \bar{q}^{\prime \prime} \rightarrow W_H(Z_H)
q_-^{\prime} + \bar{q}$ are dominated by their $q_-$ resonance
effects. The QCD NLO corrections by adopting these two subtraction
schemes enhance the LO cross section with a K-factor in the range of
$1.00 \sim 1.43$. We conclude that for associated production
processes like the $pp \rightarrow W_H q_- + X$ and $pp \rightarrow
Z_H q_- + X$ processes investigated in this paper, it is crucial to
implement a consistent and reliable on-shell subtraction scheme
separating associated production from QCD mediated pair production
properly. The scheme (II) subtracts some genuine QCD NLO
contributions, since all the $gg$ and $q\bar{q}$ initiated
contributions are excluded from the QCD NLO corrections. Therefore,
The PROSPINO scheme is more consistent that the subtraction scheme
(II).

\vskip 5mm
\par
\noindent{\large\bf Acknowledgments:} This work was supported in
part by the National Natural Science Foundation of China (Contract
No.10875112, No.11075150, No.11005101), and the Specialized Research
Fund for the Doctoral Program of Higher Education (Contract
No.20093402110030).

\end{document}